%% file: main.tex
\documentclass[lettersize,journal]{IEEEtran}
\usepackage{amsmath,amssymb,amsfonts,amsthm}
\usepackage{algorithmic}
\usepackage{algorithm}
\usepackage{array}
\usepackage{arydshln}
\usepackage[caption=false,font=normalsize,labelfont=sf,textfont=sf]{subfig}
\usepackage{textcomp}
\usepackage{stfloats}
\usepackage{url}
\usepackage{color}
\usepackage{verbatim}
\usepackage{graphicx}
\usepackage{cite}

\usepackage{hyperref}
\usepackage[normalem]{ulem}
\usepackage{multirow}
\usepackage{enumitem}
\usepackage{float}
\usepackage{caption}
\usepackage{subcaption}
\usepackage{makecell}

\usepackage{booktabs}       
\usepackage{threeparttable} 
\usepackage{amssymb}        
\usepackage{pifont}         

\usepackage{listings}
\usepackage{bm}
\usepackage{colortbl}
\usepackage{xcolor}
\usepackage{bbding}
\usepackage{hhline}
\usepackage{rotating}
\usepackage{diagbox}

\usepackage{xspace}
\newcommand{\etal}{et al.\xspace}
\newcommand{\eg}{e.g.,\xspace}
\newcommand{\ie}{i.e.,\xspace}

\newcommand{\gray}[1]{\textcolor{gray}{#1}}

\newcommand{\model}{\emph{Traffic-MoE}\xspace}

\begin{document}

%
\title{Traffic-MoE: A Sparse Foundation Model for Network Traffic Security Analysis}

\author{Jiajun Zhou, Changhui Sun, Wentao Fu, Meng Shen,~\IEEEmembership{Member,~IEEE}, Shanqing Yu, and Qi Xuan,~\IEEEmembership{Senior Member,~IEEE}

\IEEEcompsocitemizethanks{
\IEEEcompsocthanksitem This work was supported in part by National Natural Science Foundation of China (No. 62503423), in part by the National Key Research and Development Program of China (No. 2025YFA1510900), in part by the Key Research and Development Program of Zhejiang Province (No. 2026C02A1233), in part by the Yangtze River Delta Science and Technology Innovation Community Joint Research Project (No. 2026ZY03003, No. 2025CSJGG01000). (Corresponding author: Jiajun Zhou.)
\IEEEcompsocthanksitem Jiajun Zhou, Shanqing Yu, Qi Xuan, Changhui Sun, Wentao Fu are with the Institute of Cyberspace Security, Zhejiang University of Technology, Hangzhou 310023, China, with the Binjiang Cyberspace Security Institute of ZJUT, Hangzhou, 310056, China, and with the Soovar Technologies Co., Ltd., Hangzhou 310056, China (e-mail: jjzhou@zjut.edu.cn).
\IEEEcompsocthanksitem Meng Shen is with the School of Cyberspace Science and Technology, Beijing Institute of Technology, Beijing 100081, China (e-mail: shenmeng@bit.edu.cn). 
}
}



\maketitle
\begin{abstract}
As adversaries increasingly weaponize encryption and protocol obfuscation to evade traffic detection, traditional methods are rendered obsolete, necessitating deep learning to unmask sophisticated threats. However, the prohibitive computational costs of existing large models create a critical defense gap, hindering their deployment in real-time and throughput-sensitive environments. To close this vulnerability, we introduce \model, a sparse foundation model tailored for traffic security analysis. By dynamically routing traffic tokens to a small subset of specialized experts, \model effectively decouples model capacity from computational overhead. Extensive evaluations across four security-oriented tasks demonstrate that \model achieves state-of-the-art or highly competitive performance compared to leading competitors. Crucially, it delivers a 70.42\% increase in throughput, reduces inference latency by 41.39\% while significantly optimizing GPU memory consumption. Beyond efficiency, \model exhibits superior robustness against adversarial traffic shaping and maintains strong detection capabilities in few-shot scenarios, establishing a scalable and resilient paradigm for modern network traffic security analysis.
\end{abstract}


\section{Introduction}
Network communication has rapidly evolved with the widespread adoption of encryption, heterogeneous device ecosystems, and emerging paradigms such as IoT and Web3~\cite{QUIC-transport-protocol,IoT-security}. Today, TLS-based services, device-to-cloud telemetry, and decentralized protocols dominate Internet traffic, obscuring application semantics from traditional packet inspection~\cite{encrypted-traffic-analysis}. While these advancements improve privacy and availability, they also pose a critical security challenge: malicious activities, such as botnet communications and credential misuse, can hide within encrypted traffic and remain indistinguishable from benign services at the packet level~\cite{cisco-identifying, Robust-detect}.

Traffic analysis is fundamental to network defense, supporting tasks such as intrusion detection \cite{Deep-Full-Range}, anomaly detection \cite{anomaly-detection} and service attribution \cite{Qos}. However, classical machine-learning approaches rely on handcrafted statistical features or protocol signatures, which are fragile under encryption, obfuscation, and protocol mimicry. To address this, recent work leverages deep learning models that operate directly on raw byte sequences or image-based representations, achieving higher accuracy on heterogeneous and encrypted traffic \cite{traffic-classification,FS-Net}. Following trends in computer vision and natural language processing, state-of-the-art (SOTA) pre-trained traffic analysis models increasingly leverage large transformer-based architectures to further improve downstream performance \cite{ET-BERT,NetGPT,TrafficFormer}. 

Despite these advances, most models employ \textit{dense} architectures that activate all parameters during inference, causing computational cost to scale with model size. As model capacity grows, inference latency becomes prohibitively high, inevitably inducing detection blind spots and creating a critical bottleneck for real-time network traffic security analysis.
Although pre-trained models capture rich protocol semantics, their dense architectures force a trade-off between detection depth and system availability: large models may perform well on offline benchmarks yet struggle in operational environments where high throughput and low latency are critical. A practical defense system therefore requires models that preserve semantic expressiveness while enabling efficient inference.

We address this security-efficiency gap by rethinking how large-scale traffic analysis models should allocate computation. We observe that network traffic behaviors are structurally heterogeneous rather than uniformly distributed, as malware beacons, VPN tunnels, Web3 relay traffic, and benign application flows all exhibit distinct byte-level patterns and temporal dynamics \cite{Robust-detect}. 
A uniform dense architecture inefficiently forces the entire network to process all patterns, whereas a specialized-expert model can dynamically activate only the parameter subspaces most relevant to the traffic domain being processed~\cite{MOE}.
This leads to our central insight: 
\emph{Efficiency in network traffic modeling does not necessarily stem from shrinking models, but rather from selectively activating model capacity during inference} \cite{Mixtral8x7B}.

In this work, we propose \model, a sparse Mixture-of-Experts (MoE) \cite{MOE} foundation model for network traffic analysis. Instead of activating all parameters during inference, \model routes each token through a small subset of specialized experts, decoupling total detection capability from computational overhead. This design enables scaling the model parameter count by orders of magnitude without proportionally increasing inference cost, making large-scale pre-trained traffic analysis models feasible for latency-sensitive scenarios. To support heterogeneous network flows, we further develop \textit{Traffic2Token}, a unified byte-level representation method combining protocol metadata with selective payload segments, facilitating the unified modeling of cross-layer traffic patterns.
\model is trained using autoregressive self-supervision over massive unlabeled traffic corpora, capturing protocol semantics, flow directionality, and temporal dependencies. During downstream adaptation, \model can be fine-tuned for intrusion detection, service classification, VPN/Tor analysis, and other security tasks while preserving real-time inference performance, making it suitable not only for benchmarking but for operational deployment.
The main contributions of this work are summarized as follows:
\begin{itemize}[leftmargin=10pt]
    \item \textbf{A sparse foundation model for network traffic security analysis.} We propose \model, the first MoE architecture tailored to traffic modeling, enabling large model capacity while maintaining real-time inference, bridging the efficiency gap inherent in dense pre-trained models.
    \item \textbf{A unified learning framework enabling expert specialization across traffic semantics.} We support heterogeneous network flows through Traffic2Token and balanced sparse routing, allowing experts to specialize in distinct traffic behaviors and improving security-oriented generalization.
    \item \textbf{Comprehensive evaluations demonstrate the superiority and efficiency.} Extensive experiments validate that \model outperforms SOTA competitors, achieving a 70.42\% throughput increase and a 41.39\% latency reduction while optimizing memory utilization. It also exhibits robust performance in few-shot scenarios and under distribution shifts.
    \end{itemize}

\section{Related Work} \label{sec:related work}
\subsection{Statistical Machine Learning Methods}
Early traffic analysis methods rely on Deep Packet Inspection (DPI), matching predefined signatures within packet headers or payloads~\cite{DPI}. While effective in unencrypted scenarios, DPI incurs high computational overhead and becomes unreliable as modern protocols increasingly employ end-to-end encryption and payload obfuscation. 

To overcome payload visibility reliance, statistical feature-based machine learning (ML) approaches extract handcrafted features for classification. Moore \etal~\cite{moore2005toward} reduce dependence on port-based heuristics by focusing on payload features, while Saber \etal\cite{saber2018encrypted} combine PCA with SVM to model time-based statistical patterns. AppScanner~\cite{AppScanner} trains a Random Forest on 54 statistical features to identify mobile applications.
FlowPrint~\cite{Flowprint} leverages clustering to capture temporal correlations among network flows and generates application-specific traffic fingerprints, enabling the identification of previously unseen applications. Although computationally efficient, these methods depend heavily on feature engineering and struggle to capture deep semantic patterns, rendering them inadequate against encrypted or adversarial traffic conditions.

\subsection{Deep Learning Methods}
Following the rise of deep learning (DL), end-to-end models have been widely adopted to automatically extract features from raw traffic data, reducing reliance on handcrafted features. These approaches handle traffic using various data representations and model architectures. For sequence modeling, FS-Net~\cite{FS-Net} employs a bidirectional GRU-based encoder-decoder architecture to directly learn features from encrypted traffic flows. For image-based representations, ATVITSC~\cite{ATVITSC} transforms payload bytes into images and combines vision transformers with LSTM modules to jointly capture spatial and temporal dependencies. Graph-based models further treat flows as structured relational data. TFE-GNN \cite{TFE-GNN} encodes packet headers and payloads as graph nodes, while Zhou \etal~\cite{FlowID} introduce hypergraph structure that capture higher-order correlations through KNN-derived hyperedges.
These methods demonstrate the ability of deep neural networks to learn semantic structure beyond handcrafted features. However, they remain heavily dependent on large-scale labeled datasets, which are expensive to construct in real-world environments \cite{machine-survey}. Label scarcity is especially pronounced in critical long-tail classes such as zero-day attacks and malware variants, where samples are sparse and annotation requires expert knowledge. As a result, even advanced DL approaches experience significantly degraded performance when deployed in practical scenarios with limited labeled data.

\subsection{Large Pre-trained Methods}
Recent studies introduce pre-training paradigm into traffic analysis. Models such as ET-BERT~\cite{ET-BERT}, NetGPT~\cite{NetGPT}, and  TrafficFormer~\cite{TrafficFormer} perform self-supervised learning on large-scale traffic corpora to acquire generalizable representations. To address the heavy computational overhead of standard Transformers, recent efforts like YaTC~\cite{YaTC} and NetMamba~\cite{netmamba} have explored efficiency optimizations, with the latter leveraging Mamba for faster inference.
These methods significantly improve performance on network traffic detection under limited supervision, demonstrating the promise of the pre-train and fine-tune paradigm. However, these models generally remain dense, where all parameters are activated during inference. As noted in~\cite{MOE}, model capacity and inference cost remain tightly coupled, leading to resource bottlenecks that limit deployment in real-time operational networks.

\section{Threat Model}
\begin{figure}[!t]
    \centering
    \includegraphics[width=0.9\linewidth]{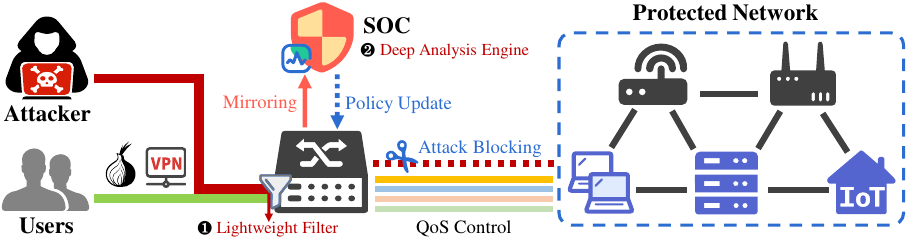}
    \caption{Threat model and operational scenario.}
    \label{fig: threat-model}
\end{figure}

\textbf{Deployment Scenario.}
Figure~\ref{fig: threat-model} illustrates a hierarchical monitoring architecture that deploys a two-stage filter at the network edge to balance throughput and inspection depth. A primary lightweight filter first intercepts plain threats~\cite{sp-ids}, while suspicious or encrypted flows are mirrored for deep out-of-path inspection. This design ensures minimal latency overhead and triggers real-time policy updates (\eg blocking or QoS adjustment)~\cite{sdn-security}. Given massive backbone traffic, the analysis engine must maximize parameter efficiency to sustain low-latency processing without detection blind spots.

\textbf{Adversary Capabilities.} We assume a sophisticated adversary aiming to disrupt services or exfiltrate data while evading detection. The adversary employs encryption and anonymization to conceal payload contents and communication endpoints~\cite{2025-openvpn}. 
To defeat statistical analysis, adversaries employ traffic shaping techniques to distort deterministic flow fingerprints~\cite{malware-tls,dpi-evasion-attack}. Furthermore, they actively exploit distribution shift strategies to launch zero-day attacks or generate out-of-distribution (OOD) traffic, attempting to bypass models trained on stationary datasets.

\textbf{Knowledge and Constraints.}
Under a black-box assumption, adversaries are cognizant of the presence of traffic analysis systems yet remain agnostic to the specific defense architecture, lacking white-box access to model parameters or gradients~\cite{sok-sp}. Furthermore, attack strategies are subject to strict functional constraints, \eg traffic manipulation must neither compromise the reliability of the command-and-control (C2) channel nor render communication indistinguishable from random noise. This is because perfect mimicry typically incurs unacceptable operational costs~\cite{machine-survey,Tantra}.

\textbf{Defender Capabilities.}
While the defender possesses full visibility of mirrored traffic headers and encrypted payloads, operations are conducted under strict constraints. Specifically, due to privacy regulations and protocols implementing Perfect Forward Secrecy (PFS), payloads cannot be decrypted, compelling the analysis engine to rely exclusively on plaintext metadata, temporal sequences, and underlying behavioral patterns~\cite{WF-ccs,packet-len-detect}. Furthermore, we assume a trusted infrastructure with a secure traffic collection and analysis pipeline, rendering threats such as physical compromise of the SOC or pre-training data poisoning out of scope.

\begin{figure*}
    \centering
    \includegraphics[width=\textwidth]{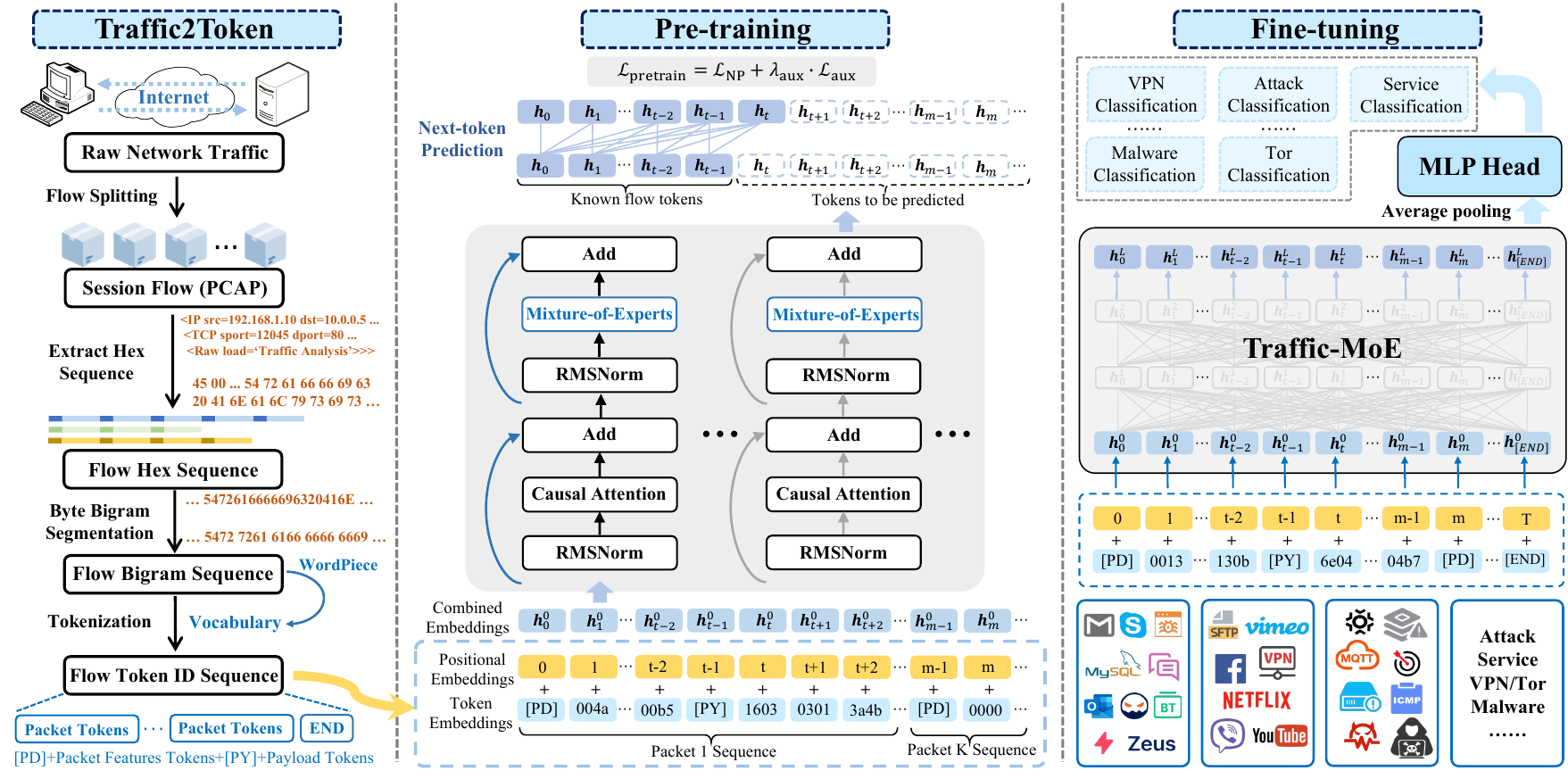}
    \caption{The overall framework of \model. It consists of the Traffic2Token serialization module, the sparse Mixture-of-Experts (MoE) backbone, and the pre-training/fine-tuning pipeline for downstream security tasks.}
    \label{fig: framework}
\end{figure*}

\section{Traffic-MoE Architecture}
Network traffic exhibits heterogeneous behavioral patterns that necessitate domain-specific modeling rather than uniform dense computation. Meanwhile, real-world security monitoring systems require low-latency processing to prevent packet drops, making dense large-scale models impractical. These challenges create a deployment gap: semantic modeling of network traffic is increasingly effective but not computationally feasible in operational environments. 
To address these constraints, \model is designed as a sparse foundation model for network traffic analysis that balances semantic expressiveness with computational efficiency. Instead of shrinking model capacity, \model activates only a subset of model parameters per input, enabling scalability without proportional inference overhead. This sparse activation strategy adapts the computational pathway to specific traffic semantics, enabling domain specialization across heterogeneous flows.

Figure~\ref{fig: framework} presents the overall workflow. 
Specifically, \model consists of four components: \ding{182} Traffic2Token converts raw flows into structured byte-level token sequences that preserve critical metadata and structural patterns; \ding{183} A hierarchical transformer backbone models temporal dependencies across packets to obtain contextualized embeddings; \ding{184} Sparse MoE layers replace dense feed-forward layers and activate only top-$k$ experts based on token semantics, combining shared experts with domain-specialized ones; \ding{185} A pretrain-finetune pipeline enables large-scale self-supervised learning and downstream task adaptation, supporting scenarios such as service classification and malicious traffic detection.
Together, these components form a unified architecture that preserves semantic modeling capability while ensuring high efficiency in throughput-sensitive environments.

\begin{figure}[!t]
    \centering
    \includegraphics[width=\linewidth]{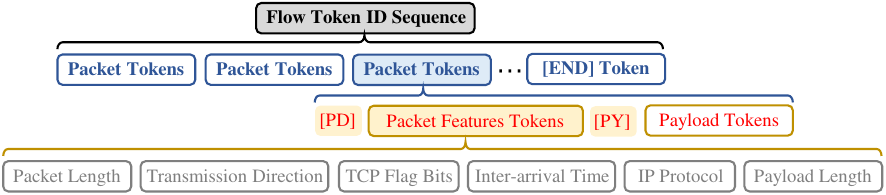}
    \caption{Illustration of flow token sequence representation.}
    \label{fig: token-seq}
\end{figure}

\subsection{Traffic Tokenization}
Raw traffic collected from operational networks contains heterogeneous communication patterns across applications, protocols, and devices. Directly modeling raw packets is ineffective as full headers introduce substantial noise while encrypted payloads limit semantic visibility. 
To bridge this gap, we design a Traffic2Token module that converts raw traffic into structured token sequences, preserving byte-level behavioral signals without protocol-specific assumptions.

\subsubsection{Packet-Based Flow Serialization}
We first reconstruct session-level flows from packet traces using five-tuple identifiers (IPs, ports, and protocol) establishing the basic processing unit for Traffic2Token.
To minimize noise from randomized or transport-irrelevant header fields, each packet is transformed into a compact and stable representation. Specifically, we extract six packet-level attributes reflecting transport and timing dynamics: packet length, direction, TCP flags, inter-arrival time, protocol type, and payload length, as shown in Figure~\ref{fig: token-seq}. These attributes are serialized into a fixed-length byte sequence. In parallel, we sample the first $J$ payload bytes to capture representative structure patterns while controlling sequence length. The metadata bytes and sampled payload bytes are then concatenated to form the packet-level hex sequence.
For each flow, the hex representations of the first $K$ packets are chronologically concatenated to construct a flow-level hex sequence. This packet-based serialization preserves communication directionality, burstiness, and timing structure.

\subsubsection{Tokenization and Embedding}
Network protocols often exhibit short-range structural regularities such as protocol-specific header fields, framing patterns, and data boundaries. To expose such correlations, we apply a sliding window to the flow hex sequence to generate byte bigrams. This transformation yields a fine-grained sequence highlighting local structural cues without full header retention.
We then treat the bigram sequences from all flows as a corpus and employ the WordPiece \cite{wordpiece} algorithm to construct a sub-byte-level vocabulary $\mathcal{V}$. WordPiece adaptively identifies statistically significant bigram combinations and assigns them as token units. This data-driven vocabulary captures recurrent traffic patterns while avoiding vocabulary explosion.
Furthermore, to enhance the model's perception of the session flow's hierarchical structure, we introduce function-specific markers into the sequence: [PD] for the beginning of each packet sequence, [PY] separating metadata fields from payload bytes, [PAD] for sequence padding, [END] denoting flow termination, and [UNK] representing low-frequency out-of-vocabulary byte patterns.
These structural markers ensure that packet boundaries remain clearly identifiable to the model when processing long sequences, preventing the semantic collapse of flow information within a flattened byte stream.
Ultimately, the bigram tokens and special markers are mapped to Token IDs via the learned vocabulary.
Each Token ID is embedded into a dense vector and summed with a positional encoding, producing the flow token sequence representation: $\boldsymbol{X}=\{\boldsymbol{x}_1,\boldsymbol{x}_2 ,\dots,\boldsymbol{x}_T\}$, where $\boldsymbol{x}_t \in \mathbb{R}^d$ is the initial token embedding in position $t$, $T$ is the sequence length, and $d$ is the embedding dimension. 
This unified embedding supports both local byte-level pattern modeling and global packet-level temporal reasoning, enabling robust pre-training and sparse expert routing.

\subsection{Backbone with Sparse Expert Layers}
\begin{figure}[!t]
    \centering
    \includegraphics[width=\linewidth]{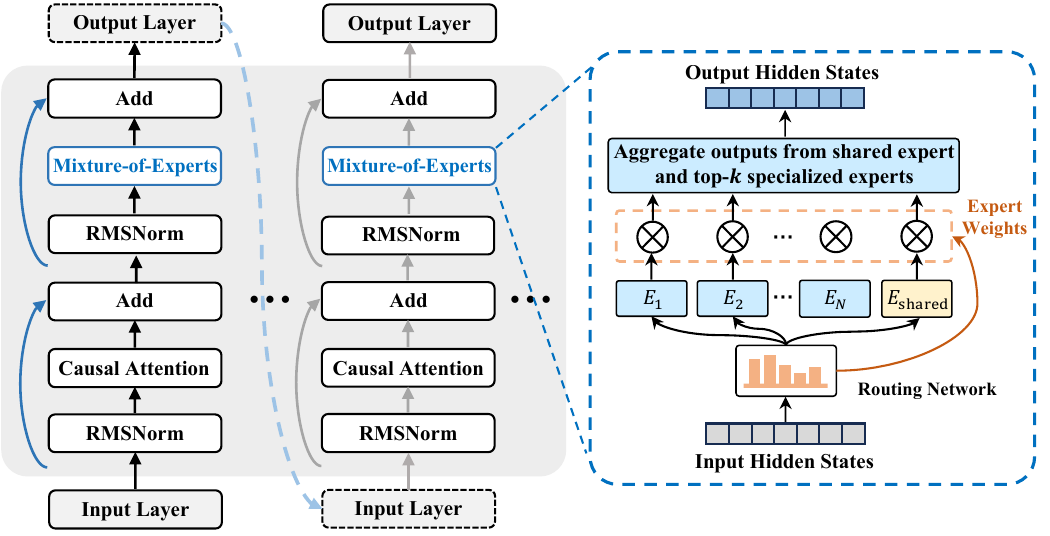}
    \caption{Illustration of \model backbone architecture.}
    \label{fig: backbone}
\end{figure}

Figure~\ref{fig: backbone} illustrates the backbone design of \model, which interleaves causal masked self-attention with sparse MoE layers to address the dual nature of network traffic. This architecture is designed to capture two distinct dimensions of traffic semantics: universal protocol syntax and heterogeneous application behaviors.
In the sequence dimension, dense self-attention operates globally across tokens to model the shared state transitions and temporal dependencies inherent in the TCP/IP protocol suite. This builds a robust contextual foundation of the underlying ``protocol grammar" (\eg handshake). In the token dimension, sparse MoE layers independently process individual tokens via dynamic routing to decouple and extract high-order, domain-specific semantic features. By stacking these layers, the backbone achieves a progressive, layer-wise abstraction, transitioning from global syntax aggregation to localized feature purification. This collaborative design, supplemented by shared expert anchors, ensures that \model can achieve deep specialization for complex downstream tasks without losing its grasp on fundamental protocol rules, providing a comprehensive and hierarchical representation of network flows.

\subsubsection{Causal Masked Self-Attention}
Network flows evolve through ordered protocol phases such as handshake exchanges, tunneling encapsulation, retransmissions, and periodic keep-alives, which manifest as temporal structures in their byte-level token sequences. To model these sequential dependencies while respecting real-world monitoring constraints where future packets are unavailable at inference time, \model employs a causal masked multi-head self-attention mechanism.

Given hidden states $\boldsymbol{H}^{(l-1)} \in \mathbb{R}^{T \times d}$ from the $(l-1)$-th block ($\boldsymbol{H}^0=\boldsymbol{X}$), we first apply Root Mean Square Layer Normalization (RMSNorm) \cite{RMSNorm} to stabilize feature scales across heterogeneous flow segments and improve training stability. 
Subsequently, to encode temporal offsets between packets, such as burst intervals or phase transitions, we incorporate Rotary Positional Embeddings (RoPE) \cite{rope} into the query and key projections. Unlike absolute positional encodings, RoPE mathematically injects relative position information by rotating the feature vectors in a shared frequency space. The query ($\boldsymbol{Q}$), key ($\boldsymbol{K}$), and value ($\boldsymbol{V}$) matrices for the $j$-th attention head are computed as:
\begin{equation}
    \label{eq: QKV}
    \begin{aligned}
    \boldsymbol{Q}_j^{(l)} &= \operatorname{RoPE}\left(\operatorname{RMSNorm}\left(\boldsymbol{H}^{(l-1)}\right)\cdot\boldsymbol{W}_{Q,j}^{(l)}\right), \\
    \boldsymbol{K}_j^{(l)} &= \operatorname{RoPE}\left(\operatorname{RMSNorm}\left(\boldsymbol{H}^{(l-1)}\right)\cdot\boldsymbol{W}_{K,j}^{(l)}\right), \\
    \boldsymbol{V}_j^{(l)} &= \operatorname{RMSNorm}\left(\boldsymbol{H}^{(l-1)}\right)\cdot\boldsymbol{W}_{V,j}^{(l)}\ ,
    \end{aligned}
\end{equation}
where $\boldsymbol{W}_{\{Q,K,V\},j}^{(l)} \in \mathbb{R}^{d \times d_m}$ are learnable projection matrices. This formulation enables the attention mechanism to capture relative sequential distance between flow tokens, identifying traffic patterns invariant to absolute sequence positions (\eg attack signatures appearing at varying offsets).

To ensure traffic causality where future packets cannot influence current state, we apply a strictly lower-triangular causal mask $\boldsymbol{M} \in \mathbb{R}^{T \times T}$. The attention output is:

\begin{equation}
    \label{eq:attention}
    \begin{split}
        \boldsymbol{O}_j^{(l)} &= \operatorname{Softmax}\left(\frac{\boldsymbol{Q}_j^{(l)} (\boldsymbol{K}_j^{(l)})^\top}{\sqrt{d_m}} + \boldsymbol{M}\right)\cdot \boldsymbol{V}_j^{(l)} \\
        \boldsymbol{M}_{t p} &= \begin{cases}0, & p \leq t \\ -\infty, & p>t \end{cases}
    \end{split}
\end{equation}

This causal mask ensures that the token at position $t$ can only attend to information from itself and preceding positions.
Finally, the outputs from all $m$ attention heads are concatenated and projected to form the residual update:
\begin{equation}\label{eq: res-1}
    \widetilde{\boldsymbol{H}}^{(l)} = \boldsymbol{H}^{(l-1)} + \left[\boldsymbol{O}_1^{(l)} \  \Vert \  \boldsymbol{O}_2^{(l)} \  \Vert \    \cdots \  \Vert  \  \boldsymbol{O}_m^{(l)}\right]\cdot\boldsymbol{W}_O^{(l)},
\end{equation}
where $\boldsymbol{W}_O^{(l)} \in \mathbb{R}^{d \times d}$ is the output projection matrix, and $\widetilde{\boldsymbol{H}}^{(l)}$ is the intermediate hidden state.
This design enables autoregressive learning of traffic semantics, effectively modeling the state transition logic of network protocols.

\subsubsection{Mixture-of-Experts Layer with Sparse Routing}
Following the causal masked self-attention module, the intermediate hidden state $\widetilde{\boldsymbol{H}}^{(l)}$ is processed by a sparse MoE layer. 
Standard Transformers typically employ a dense Feed-Forward Network (FFN) that applies identical parameters to all tokens.
However, network traffic combines universal protocol regularities (\eg transport headers, handshake patterns) with highly heterogeneous behaviors (\eg application-specific traffic fingerprints). To accommodate this heterogeneity without incurring a proportional increase in computational cost, \model replaces the dense FFN with a hybrid MoE layer composed of a shared expert and multiple specialized experts.

To decouple general protocol knowledge from specific traffic behaviors, we design a dual-pathway computation. First, the input $\widetilde{\boldsymbol{H}}^{(l)}$ undergoes normalization via RMSNorm to produce $\boldsymbol{Z}^{(l)} = \operatorname{RMSNorm}(\widetilde{\boldsymbol{H}}^{(l)})$.
This normalized representation is then routed to two distinct expert groups, as defined below:

\emph{\textbf{Shared Expert for Universal Protocol Semantics.}}
To capture invariant traffic structures distinct from application-specific payloads, we employ a persistently activated \textit{shared expert}, $E_{\operatorname{shared}}$. Unlike the competitive routing of specialized experts, the shared expert utilizes an independent gate to adaptively regulate the injection of general traffic knowledge:
\begin{equation}
    \label{eq: shared expert}
    \boldsymbol{O}_{\operatorname{shared}}^{(l)} = \operatorname{Sigmoid}\left(\boldsymbol{Z}^{(l)}\boldsymbol{w}_\vartriangle\right) \odot E_{\operatorname{shared}}\left(\boldsymbol{Z}^{(l)}\right),
\end{equation}
where $\boldsymbol{w}_\vartriangle  \in \mathbb{R}^{d}$ is a projection vector, and $\odot$ denotes element-wise multiplication broadcasted over the feature dimension.

\emph{\textbf{Specialized Experts for Heterogeneous Patterns.}} 
Simultaneously, $N$ specialized experts $\{E_e\}_{e=1}^N$ are established to handle diverse traffic patterns. A routing network computes a probability distribution over these experts, selecting the Top-$k$ relevant experts to process each token. The routing scores $\boldsymbol{S}^{(l)}$ and the sparse routing weights $\widetilde{\boldsymbol{S}}^{(l)}$ are computed as:
\begin{equation}
    \label{eq: routing}
    \boldsymbol{S}^{(l)} =\operatorname{Softmax}\left(\boldsymbol{Z}^{(l)}\boldsymbol{W}_\ast \right) , \
    \widetilde{\boldsymbol{S}}_{ue}^{(l)}= \begin{cases} \boldsymbol{S}_{ue}^{(l)}, & e \in \operatorname{Top}_k(\boldsymbol{S}_\textit{u}^{(l)}) \\ 0, & \text {otherwise}\end{cases}
\end{equation}
where $\boldsymbol{W}_\ast \in \mathbb{R}^{d \times N}$ projects the input into the expert embedding space, 
$\boldsymbol{S}^{(l)} \in \mathbb{R}^{T \times N}$ contains per-token logits over experts, and $\widetilde{\boldsymbol{S}}^{(l)} \in \mathbb{R}^{T \times N}$ is a sparse routing matrix in which only the top-$k$ entries per token (per row) are retained and the remaining entries are set to zero. This sparse routing mechanism ensures that, for each token, only a small number of experts are activated according to the current traffic semantics.
The output of the sparse branch is the weighted sum of the activated experts:
\begin{equation}
    \label{eq: specialized expert}
    \boldsymbol{O}_{\operatorname{special}}^{(l)}
    = \sum_{e=1}^{N} \widetilde{\boldsymbol{S}}^{(l)}_{:,e} \odot E_e\left(\boldsymbol{Z}^{(l)}\right),
\end{equation}
where $\widetilde{\boldsymbol{S}}^{(l)}_{:,e}$ denotes routing weights for expert $E_e$ across all tokens and multiplication $\odot$ is broadcast along the feature dimension.
By setting $k \ll N$ (\eg $k=2$ in this work), \model selectively activates parameter subspaces relevant to the specific attack or protocol phase currently being processed.

For the implementation of the experts, both the shared expert and the specialized experts follow a SwiGLU-based gated feed-forward design. Given an input $\boldsymbol{Z}$, an expert parameterized by $\{\boldsymbol{W}_{\text{gate}},\boldsymbol{W}_{\text{up}},\boldsymbol{W}_{\text{down}}\}$ computes:
\begin{equation}
    \label{eq: expert}
    E(\boldsymbol{Z}) 
    = 
    \left(
        \operatorname{SiLU}\left(\boldsymbol{Z}\boldsymbol{W}_{\mathrm{gate}}\right)
        \odot 
        \left(\boldsymbol{Z}\boldsymbol{W}_{\mathrm{up}}\right)
    \right)
    \boldsymbol{W}_{\mathrm{down}}\ ,
\end{equation}
where $\operatorname{SiLU}(\cdot)$ is the Sigmoid Linear Unit. Note that the shared expert retains a full intermediate dimension $d'\gg d$ for capturing universal protocol patterns (\ie $\boldsymbol{W}_{\{\text{gate},\text{up}\}} \in \mathbb{R}^{d \times d'}$ and $\boldsymbol{W}_{\text{down}} \in \mathbb{R}^{d' \times d}$), whereas specialized experts use a reduced intermediate dimension ($d'/k$) to balance the number of experts and computation (\ie $\boldsymbol{W}_{\{\text{gate},\text{up}\}} \in \mathbb{R}^{d \times \frac{d'}{k}}$ and $\boldsymbol{W}_{\text{down}} \in \mathbb{R}^{\frac{d'}{k} \times d}$). The weights of different specialized experts are mutually independent.

The output of the MoE layer integrates protocol-invariant features from the shared expert and instance-specific features from the specialized experts, added to the residual stream:
\begin{equation}\label{eq: res-2}
    \boldsymbol{H}^{(l)} = \widetilde{\boldsymbol{H}}^{(l)} + \left(\boldsymbol{O}_{\operatorname{shared}}^{(l)} + \boldsymbol{O}_{{\operatorname{special}}}^{(l)}\right).
\end{equation}

To prevent routing collapse, where the routing network converges to utilizing only a few experts, leaving others under-trained, we incorporate an auxiliary \textbf{load-balancing loss}, $\mathcal{L}_{\operatorname{aux}}$. For a given layer $l$, we compute the loss based on the batch-wise statistics of expert activation. Let $\boldsymbol{Load}_e^{(l)}$ be the fraction of tokens in the batch assigned to expert $e$, and $\boldsymbol{Prob}_e^{(l)}$ be the average routing probability for expert $e$ across the batch. The load-balancing loss for layer $l$ is:
\begin{equation}\label{eq: load-balance}
\mathcal{L}_{\text{aux}}^{(l)} = N \sum_{e=1}^{N} \boldsymbol{Load}_e^{(l)} \cdot \boldsymbol{Prob}_e^{(l)}\ ,\ \mathcal{L}_{\text{aux}} = \frac{1}{L}\sum_{l=1}^{L} \mathcal{L}_{\text{aux}}^{(l)}\ .
\end{equation}

The total auxiliary loss $\mathcal{L}_{\text{aux}}$ is the average of these penalties across all $L$ layers. This constraint ensures a uniform batch-level expert activation distribution, maximizing model's capacity utilization while maintaining sparse inference efficiency.

\subsection{Pre-training and Fine-tuning}
To endow \model with both broad traffic understanding and task-specific discrimination, we adopt a two-stage training framework consisting of large-scale self-supervised pre-training followed by lightweight few-shot fine-tuning. 
The pre-training stage exposes the backbone to massive unlabeled traffic data, enabling it to learn protocol-aware and temporally structured representations through autoregressive Next Token Prediction (NTP) and balanced sparse-expert activation. Fine-tuning then adapts these general representations to downstream traffic security applications by employing targeted data augmentation, flow-level semantic pooling, and hierarchical optimization. This two-stage framework enables \model to scale in capacity while remaining efficient and robust across diverse traffic security scenarios.

\subsubsection{Autoregressive Pre-training: Next Token Prediction}
Real-world networks generate massive unlabeled packet flows but only limited annotated security data, making pre-training essential for learning generalizable traffic representations. We adopt a self-supervised autoregressive NTP objective to exploit this abundant unlabeled traffic. As network communication is sequential and governed by protocol-driven state transitions, predicting the subsequent token from historical context enables the model to learn temporal dependencies (\eg TCP handshake timing) and structural regularities (\eg TLS record header formats) even in encrypted flows. This process equips \model with a unified understanding of network behaviors that can be effectively adapted during downstream fine-tuning. Figure~\ref{fig: framework} illustrates the pre-training pipeline, highlighting interaction among Traffic2Token representation, causal attention, and sparse experts to form a scalable traffic foundation model.

Given a flow token sequence $X = \{x_1, x_2, \cdots, x_T\}$, \model learns to maximize the likelihood of the next token
$x_t$ conditioned on its historical context $\boldsymbol{X}_{<t}=\{\boldsymbol{x}_{1},\boldsymbol{x}_{2},\cdots,\boldsymbol{x}_{t-1}\}$. The prefix $\boldsymbol{X}_{<t}$ is encoded through $L$-layer \model backbone to produce hidden states $\boldsymbol{H}_{<t}^{(L)}=\{\boldsymbol{h}_{1}^{(L)},\cdots,\boldsymbol{h}_{t-1}^{(L)}\}$. The final state $\boldsymbol{h}_{t-1}^{(L)} \in \mathbb{R}^d$ summarizes all contextual information up to position $t-1$ and is projected into the vocabulary space to obtain the probability distribution of next-token prediction:
\begin{equation}
    p_\theta(x_t \mid \boldsymbol{X}_{<t}) = \operatorname{Softmax}\left(\boldsymbol{h}_{t-1}^{(L)} \cdot \boldsymbol{W}_{\mathrm{vocab}} \right),
\end{equation}
where $\boldsymbol{W}_{\mathrm{vocab}} \in \mathbb{R}^{d \times |\mathcal{V}|}$ maps hidden states to vocabulary logits.
To optimize the model parameters $\theta$, we employ the negative log-likelihood loss function $\mathcal{L}_{\mathrm{NP}}$. For a mini-batch of $B$ flows, this is calculated as the average loss over all valid tokens in the batch via a binary mask $m_t^{(b)} \in \{0, 1\}$:  
\begin{equation}\label{eq: NTP}
\mathcal{L}_{\mathrm{NP}} = - \frac{1}{\sum_{b=1}^{B} \sum_{t=2}^{T} m_t^{(b)}} \sum_{b=1}^{B} \sum_{t=2}^{T} m_t^{(b)} \log p_\theta\left(x_t^{(b)} \mid \boldsymbol{X}_{<t}^{(b)}\right) .
\end{equation}

Finally, as MoE activates limited experts per token, balanced expert utilization is essential in pre-training. We incorporate load-balancing term $\mathcal{L}_{\mathrm{aux}}$ to prevent routing collapse and promote expert specialization across heterogeneous traffic behaviors. The overall pre-training objective is:

\begin{equation}
    \mathcal{L}_{\mathrm{pretrain}} = \mathcal{L}_{\mathrm{NP}} + \lambda_{\mathrm{aux}} \cdot \mathcal{L}_\mathrm{{aux}}\ ,
\end{equation}
where $\lambda_{\mathrm{aux}}$ controls the strength of load-balancing regularization. Minimizing this composite loss enables \model to acquire rich, protocol-aware, and temporally grounded representations of traffic flows, while simultaneously encouraging efficient and diverse expert activation. As a result, the pre-trained backbone captures both universal traffic patterns and domain-specific subtleties, providing a strong foundation for downstream tasks such as intrusion detection, service classification, and VPN/Tor traffic analysis.

\subsubsection{Few-shot Fine-tuning for Traffic Analysis} 
Downstream traffic analysis tasks such as malware detection, service classification and C2 identification typically face label scarcity and highly imbalanced class distributions. To effectively adapt \model to these data-constrained scenarios, we introduce a fine-tuning strategy tailored for few-shot learning. This strategy integrates temporal-aware data augmentation, flow-level representation aggregation, and hierarchical optimization to achieve robust performance under limited supervision.

Network sessions vary widely in duration. DNS queries may finish within milliseconds, whereas video streaming or VPN tunnels can persist for minutes or even hours. This discrepancy leads to a long-tailed flow length distribution and extremely limited samples for certain classes. 
To mitigate this imbalance while precluding data leakage, we apply Temporal Traffic Slicing independently within each partition after a rigorous session-level train/validation/test split, dividing long flows into fixed-duration sub-flows that inherit the original label.
The resulting augmented dataset increases sample diversity and enables the model to capture fine-grained temporal dynamics such as periodic beacons, idle intervals, and bursty transmissions that are often diluted in full-session representations. This augmentation maximizes the utility of sparse session-level examples and is particularly effective in few-shot scenarios. 

For supervised adaptation, we aggregate token-level representations from the final backbone layer into a unified flow-level embedding. Instead of relying on a single special token (\eg [PD] or [END]), which may overlook mid-session structural semantics, we apply mean pooling aggregation over all valid tokens (indicated by the binary mask $m_t$) and compute class probabilities through a MLP-based prediction head:
\begin{equation}\label{eq: classification prob}
    p_\theta(y \mid X)
    =
    \operatorname{Softmax}\!\left(
        \operatorname{MLP}\!\left(
            \frac{1}{\sum_{t=1}^{T} m_t} \sum_{t=1}^{T} m_t \boldsymbol{h}_t^{(L)}
        \right)
    \right).
\end{equation}

For a mini-batch of $B$ flow token sequences $\{X^{(b)}\}_{b=1}^{B}$ with labels $\{y_{b}\}_{b=1}^{B}$, the supervised objective is defined as the batch-averaged cross-entropy loss:
\begin{equation}\label{eq: class-loss}
    \mathcal{L}_{\text{TC}}
    =
    -
    \frac{1}{B}
    \sum_{b=1}^{B}
    \log
    p_\theta\!\left(y_{b} \mid X^{(b)}\right).
\end{equation}

To maintain stable expert routing and prevent expert collapse during adaptation, we retain the MoE load-balancing term $\mathcal{L}_{\mathrm{aux}}$, yielding the overall fine-tuning objective:
\begin{equation}
    \mathcal{L}_{\mathrm{finetune}}
    =
    \mathcal{L}_{\mathrm{TC}}
    +
    \lambda_{\mathrm{aux}}\cdot \mathcal{L}_{\text{aux}}\ .
\end{equation}

By minimizing this composite loss, \model can precisely adapt the general traffic knowledge learned during pre-training to downstream classification tasks while efficiently leveraging its powerful MoE architecture, thereby continuously improving its classification performance in specific traffic security applications.

Finally, to mitigate catastrophic forgetting of pre-trained knowledge, we employ a Layer-wise Learning Rate Decay (LLRD) schedule tailored to the hierarchical structure of \model. Specifically, shallower layers encoding generic packet syntax are updated conservatively with lower learning rates. In contrast, deeper layers and the classification head have higher learning rates, facilitating quick adaptation to task-specific semantics.
For a backbone with $L$ layers and base learning rate $\eta_{0}$, the learning rate for layer $l$ is:
\begin{equation}
    \eta_l=\xi^{\,L-l}\cdot\eta_{0}\ ,
\end{equation}
where $\xi < 1$ is the decay factor.  
This strategy preserves the universal traffic knowledge encoded during pre-training while enabling the deeper layers to adjust to the semantic and distributional characteristics of the downstream task.

By combining temporal slicing, flow-level semantic pooling, sparse-expert regularization, and hierarchical optimization, \model achieves high sample efficiency and robust generalization in data-constrained traffic security scenarios. Even with limited labeled examples, \model adapts effectively to new domains and evolving threats while retaining the broad traffic knowledge learned during pre-training.

\begin{table*}[!t]
    \centering
     \caption{Overview of datasets used in pre-training and fine-tuning phases}
     \label{tab:datasets_overview} 
     \resizebox{\textwidth}{!}{
        \begin{threeparttable}
    \begin{tabular}{ccccccccccc}
        \toprule
        \multirow{2}{*}{\textbf{Dataset}} & \multirow{2}{*}{\textbf{Domain}} & \textbf{Original} & \textbf{Original Scale\tnote{*}} & \multirow{2}{*}{\textbf{Key Protocols}} & \multicolumn{2}{c}{\textbf{Pre-training}} & \multicolumn{4}{c}{\textbf{Fine-tuning}} \\
        \cmidrule(lr){6-7} \cmidrule(lr){8-11}    
         & & \textbf{Classes} & \textbf{(Record)} & & Included & Size (Sessions) & Included & Downstream Task & Size (Sessions) & Classes \\
        \midrule                        
        
    USTC-TFC2016             & Malware Traffic         & 20                                & 0.7M rows                                        & FTP, SMB, HTTP, DNS, MySQL, $\cdots$                                                                           & \Checkmark      & 203.2k                & \multicolumn{1}{c}{\XSolidBrush}        & \multicolumn{1}{c}{-}               & -               & -        \\
    \addlinespace
    UNSW-NB15                & Network Intrusion       & 10                                & 2.5M rows                                        & HTTP, ICMP, FTP, SSH, DNS, $\cdots$                                                                            & \Checkmark      & 828.3k                & \multicolumn{1}{c}{\XSolidBrush}        & \multicolumn{1}{c}{-}               & -               & -        \\ 
    \addlinespace
    CICIoT2023               & IoT Security            & 34                                & 46.7M rows                                       & MQTT, HTTP, DNS, SSH, ARP, $\cdots$                                                                            & \Checkmark      & 892.5k                & \Checkmark                           & IoT Attack Detection                & 255.2k          & 34       \\ 
    \addlinespace
    CICIoMT2024              & IoMT Security     & 19                                & 8.8M rows                                        & Wi-Fi, MQTT, HTTP, DNS, SSH, $\cdots$                                                                          & \Checkmark      & 439.4k                & \Checkmark                           & IoMT Attack Detection               & 107.3k          & 19       \\ 
    \addlinespace
    ISCXVPN2016(NonVPN)      & Unencrypted Traffic     & \multirow{2}{*}{14}               & \multirow{2}{*}{298.9k sessions}                 & \multirow{2}{*}{\begin{tabular}[c]{@{}c@{}}OpenVPN, BitTorrent, SFTP, FTPS, \\HTTPS, SMTPS, POP3S, $\cdots$\end{tabular}} & \Checkmark      & 160.2k                & \Checkmark                           & NonVPN Traffic Detection            & 23.9k           & 6        \\ 
    ISCXVPN2016(Mixed)       & VPN Encrypted Traffic   &                                   &                                                  &                                                                                                                & \XSolidBrush       & -                     & \Checkmark                           & Mixed Traffic Detection             & 26.2k           & 7        \\ 
    \addlinespace
    ISCXTor2016(NonTor)      & Unencrypted Traffic     & \multirow{2}{*}{8}                & \multirow{2}{*}{53.7k sessions}                  & \multirow{2}{*}{\begin{tabular}[c]{@{}c@{}}Tor, BitTorrent, HTTPS, SMTP, \\POP3, SFTP, FTPS, $\cdots$\end{tabular}}         & \XSolidBrush       & -                     & \Checkmark                           & NonTor Traffic Detection            & 51.5k           & 8        \\ 
    ISCXTor2016(Tor)         & Tor Encrypted Traffic   &                                   &                                                  &                                                                                                                & \XSolidBrush       & -                     & \Checkmark                           & Tor Traffic Detection               & 2.0k            & 8        \\
    
    \addlinespace
    CipherSpectrum           & TLS Encrypted Traffic   & 40                                & 120.0k sessions                                  & \begin{tabular}[c]{@{}c@{}}TLS-AES-128-GCM, TLS-AES-256-GCM,\\ TLS-CHACHA20-POLY1305\end{tabular} & \XSolidBrush       & -                     & \Checkmark                           & TLS Traffic Detection               & 41.2k           & 10       \\
   
    \bottomrule      
    \end{tabular}
     \begin{tablenotes}
        \small 
        \item[*] \textbf{Note:} The unit ``rows'' refers to pre-processed data records (\eg packet windows or flow images) as defined in the original datasets, whereas ``sessions'' refers to complete bidirectional flows reconstructed from raw PCAP traces. For datasets appearing in both pre-training and fine-tuning, the sessions allocated to each phase are strictly non-overlapping.
        \end{tablenotes}
\end{threeparttable}
}
\end{table*}

\section{Experiment Evaluation}
\subsection{Datasets and Downstream Tasks}
\subsubsection{Pre-training Corpus}
To facilitate the learning of universal and robust traffic representations across heterogeneous network environments, we construct a comprehensive pre-training corpus aggregating over \textbf{2 million unlabeled flows} from five authoritative public datasets (\textit{USTC-TFC2016}~\cite{USTC-TFC2016}, \textit{UNSW-NB15}~\cite{UNSW-NB15}, \textit{CICIoT2023}~\cite{CICIoT2023}, \textit{CICIoMT2024}~\cite{CICIoMT2024}, \textit{ISCXVPN2016(NonVPN)}~\cite{ISCXVPN2016}). These datasets encompass a wide spectrum of encryption complexities, protocol diversities, and behavioral signatures. 
Unlike prior works restricted to domain-specific subsets, our corpus incorporates extensive \textit{distributional diversity} by intertwining standard IT traffic (\eg HTTP/DNS) with specialized OT protocols (\eg MQTT/Zigbee) and heavily obfuscated tunnels. This cross-domain dataset synthesis forces the model to learn universal and intrinsic traffic semantics rather than dataset-specific artifacts.

\subsubsection{Fine-tuning Datasets and Downstream Tasks}
During fine-tuning phase, we select five challenging public datasets to design four downstream tasks, systematically evaluating model's transfer learning capability from attack detection to service classification: 
\ding{182} For \textbf{Tor/NonTor Service Classification}, we leverage \textit{ISCXTor2016}\cite{ISCXTor2016} to classify 8 application services within separated ``Tor'' and ``NonTor'' traffic subsets, challenging the model to capture subtle flow-level fingerprints in extreme anonymization scenarios;
\ding{183} For \textbf{NonVPN/Mixed Service Classification}, we employ \textit{ISCXVPN2016} to construct two sub-tasks, ``NonVPN'' and ``Mixed'' service classification, assessing the restoration of service semantics under tunnel encapsulation;
\ding{184} For \textbf{IoT/IoMT Attack Detection}, we utilize \textit{CICIoT2023} to identify 33 distinct attack behaviors, and \textit{CICIoMT2024} to detect 18 targeted attacks, testing performance in sensitive healthcare environments involving protocols like MQTT.
\ding{185} For \textbf{Modern Encrypted Service Classification}, we employ \textit{CipherSpectrum}~\cite{cipherspectrum} and group domains into 10 typical service categories, validating the model’s ability to distinguish fine-grained services under pure TLS 1.3 encryption.
Detailed descriptions are provided in Table~\ref{tab:datasets_overview} and Appendix B.

\subsection{Comparison Methods and Parameter Setup}
\subsubsection{Comparison Methods}
To precisely benchmark \model, we evaluate it against eight SOTA baselines categorized into two paradigms. As core baselines representing the ``pre-training and fine-tuning'' paradigm, we select \textit{ET-BERT}~\cite{ET-BERT}, \textit{NetGPT}~\cite{NetGPT}, \textit{TrafficFormer}~\cite{TrafficFormer} \textit{YaTC}~\cite{YaTC}, and \textit{NetMamba}~\cite{netmamba} for comparison. Additionally, to verify the effectiveness of the pre-training phase, we include two ML methods (\textit{AppScanner}~\cite{AppScanner}, \textit{FlowPrint}~\cite{Flowprint}) and one DL method (\textit{FS-Net}~\cite{FS-Net}) for comparison. To ensure fairness, all comparison models are reproduced based on their official open-source code. 
Specifically, the pre-training methods utilize the exact same pre-training and fine-tuning datasets constructed in this work, adopting identical early stopping strategies during fine-tuning to ensure optimal convergence, while the ML/DL methods are trained directly on the constructed fine-tuning datasets. More details of comparison methods are presented in Appendix C.

\subsubsection{Implementation Details and Parameter Configuration}
All experiments are implemented in PyTorch 2.2 and executed on a high-performance cluster equipped with Intel Xeon Gold 5218R CPUs and NVIDIA A100 GPUs (40GB VRAM). The environment runs on Ubuntu 22.04 LTS with CUDA 11.7.

\textbf{Data Preprocessing and Serialization:} 
PCAP traces are first segmented into independent sessions via SplitCap. For datasets utilized in both pre-training and fine-tuning phases, these sessions are rigorously partitioned into two mutually exclusive subsets to ensure the complete exclusion of fine-tuning data from the pre-training corpus, prior to independent serialization via \textit{Traffic2Token}.
Specifically, we extract 11 bytes of header features and the first $J=40$ bytes of payload per packet, truncating each session to the first $K=10$ packets to balance computational efficiency and early-stage detection. 

\textbf{Training Configuration:} We utilize AdamW optimizer with a batch size of $B=32$.
\textbf{For the pre-training phase}, the model is trained for 8 epochs with a learning rate of $3.0 \times 10^{-4}$. The loss function combines the NTP loss and the auxiliary load-balancing loss, with the regularization coefficient set to $\lambda_{\text{aux}} = 0.02$.
\textbf{For the fine-tuning phase, we utilize session samples that are strictly disjoint from the pre-training corpus to prevent data leakage.} These data are split into training, validation, and testing sets with an 8:1:1 ratio. We employ a LLRD schedule to preserve pre-trained knowledge, setting the decay factor $\xi = 0.9$ and the base learning rate $\eta_0 = 5.0 \times 10^{-5}$. The maximum training duration is set to 40 epochs, safeguarded via early stopping with a patience of 5 epochs monitoring the validation Macro-F1 score.

\subsubsection{Evaluation Metrics}
Given the inherent long-tail distribution and class imbalance in network traffic, standard accuracy is insufficient for objective evaluation. We therefore adopt a multi-dimensional evaluation protocol. 
For evaluating general performance, we select Macro-Precision (M-PR), Macro-Recall (M-RC), and Macro-F1 Score (M-F1) as core metrics. By calculating performance independently for each class before averaging, these macro-metrics treat minority classes (\eg emerging threats) and majority classes equally, preventing results from being biased by high-volume benign traffic.
For evaluating security effectiveness, we specifically report the False Negative Rate (FNR) and False Positive Rate (FPR). These metrics are critical for assessing the operational risks of the model, quantifying the trade-off between missed threats and false alarms in real-world deployments.

\begin{table*}[!t]
    \centering
    \caption{Comparison with state-of-the-art methods on different traffic detection tasks}
    \label{tab:method_comparison}

    \resizebox{\textwidth}{!}{%
        \begin{tabular}{@{}c@{}}
            
            \begin{tabular}{c cccc cccc cccc cccc}
                \toprule
                \multirow{2}{*}{\textbf{Method}} & 
                \multicolumn{4}{c}{\textbf{Task 1: ISCXTor2016(NonTor)}} & 
                \multicolumn{4}{c}{\textbf{Task 1: ISCXTor2016(Tor)}} & 
                \multicolumn{4}{c}{\textbf{Task 2: ISCXVPN2016(NonVPN)}} & 
                \multicolumn{4}{c}{\textbf{Task 2: ISCXVPN2016(Mixed)}} \\
                \cmidrule(lr){2-5} \cmidrule(lr){6-9} \cmidrule(lr){10-13} \cmidrule(lr){14-17}
                 & ACC & M-PR & M-RC & M-F1 & ACC & M-PR & M-RC & M-F1 & ACC & M-PR & M-RC & M-F1 & ACC & M-PR & M-RC & M-F1 \\
                \midrule
                
                AppScanner & 0.9468 & 0.8587 & 0.6934 & 0.7436 & \underline{0.8690} & 0.7921 & 0.7054 & \underline{0.7254} & 0.5794 & 0.6756 & 0.6782 & 0.6588 & 0.6249 & 0.7414 & 0.7355 & 0.7248 \\
                FlowPrint & 0.9213 & 0.7408 & 0.7245 & 0.7138 & 0.4127 & 0.1907 & 0.2595 & 0.1738 & 0.5459 & 0.6283 & 0.5620 & 0.5617 & 0.5659 & 0.6033 & 0.6431 & 0.5952 \\
                FS-Net & 0.9406 & 0.7841 & 0.6518 & 0.6972 & 0.6679 & 0.2008 & 0.2527 & 0.2161 & 0.5093 & 0.4952 & 0.5021 & 0.4971 & 0.5357 & 0.4617 & 0.4681 & 0.4631 \\
                ET-BERT & 0.9653 & 0.8322 & 0.7706 & 0.7960 & 0.7932 & 0.4093 & 0.4677 & 0.4303 & 0.5502 & 0.6670 & 0.6852 & 0.6755 & 0.6106 & 0.7503 & 0.7549 & 0.7521 \\
                NetGPT & 0.9655 & 0.8479 & 0.7911 & 0.8134 & 0.8506 & 0.7342 & 0.6805 & 0.6583 & 0.6618 & 0.7405 & \underline{0.7511} & \underline{0.7457} & 0.6973 & 0.8003 & \underline{0.8072} & \underline{0.8034} \\
                TrafficFormer & 0.9554 & 0.7663 & 0.7596 & 0.7418 & 0.7975 & 0.6083 & 0.5984 & 0.5428 & 0.6045 & 0.6831 & 0.6970 & 0.6842 & 0.6373 & 0.7689 & 0.7285 & 0.7422 \\
                YaTC & \underline{0.9736} & \underline{0.9001} & 0.8203 & \underline{0.8502} & 0.8621 & \underline{0.8028} & \underline{0.7565} & 0.7106 & \underline{0.6860} & \underline{0.7550} & 0.7494 & 0.7207 & \underline{0.7022} & \underline{0.8036} & 0.7814 & 0.7849 \\
                NetMamba & 0.9715 & 0.8554 & \underline{0.8287} & 0.8414 & 0.7488 & 0.5550 & 0.4856 & 0.4892 & 0.6417 & 0.6904 & 0.6543 & 0.6622 & 0.6605 & 0.7117 & 0.6760 & 0.6756 \\
                \textbf{Traffic-MoE} & \textbf{0.9827} & \textbf{0.9222} & \textbf{0.8707} & \textbf{0.8900} & \textbf{0.9089} & \textbf{0.8942} & \textbf{0.7879} & \textbf{0.8072} & \textbf{0.7613} & \textbf{0.7866} & \textbf{0.8158} & \textbf{0.8005} & \textbf{0.7679} & \textbf{0.8306} & \textbf{0.8377} & \textbf{0.8332} \\
                \bottomrule
            \end{tabular}
            
            \\ \vspace{10pt} \\ 
            
            \begin{tabular}{c cccc cccc cccc}
                \toprule
                \multirow{2}{*}{\textbf{Method}} & 
                \multicolumn{4}{c}{\textbf{Task 3: CICIoMT2024}} & 
                \multicolumn{4}{c}{\textbf{Task 3: CICIoT2023}} & 
                \multicolumn{4}{c}{\textbf{Task 4: CipherSpectrum}} \\
                \cmidrule(lr){2-5} \cmidrule(lr){6-9} \cmidrule(lr){10-13}
                 & ACC & M-PR & M-RC & M-F1 & ACC & M-PR & M-RC & M-F1 & ACC & M-PR & M-RC & M-F1 \\
                \midrule
                
                AppScanner & 0.6929 & 0.7473 & 0.6783 & 0.6397 & 0.5266 & 0.5535 & 0.4471 & 0.4619 & \underline{0.9276} & \underline{0.9248} & \underline{0.9199} & \underline{0.9270} \\
                FlowPrint & 0.0132 & 0.0373 & 0.1364 & 0.0501 & 0.4578 & 0.3166 & 0.2048 & 0.1989 & 0.2900 & 0.2580 & 0.2702 & 0.2602 \\
                FS-Net & 0.5788 & 0.5319 & 0.5050 & 0.4533 & 0.4903 & 0.5265 & 0.4046 & 0.4094 & 0.8714 & 0.8447 & 0.8359 & 0.8388 \\
                ET-BERT & \textbf{0.9769} & 0.5620 & 0.5271 & 0.5255 & 0.7455 & 0.5416 & 0.5172 & 0.5217 & 0.4630 & 0.4267 & 0.3618 & 0.3474 \\
                NetGPT & 0.8958 & 0.8326 & 0.8335 & 0.8300 & 0.7684 & 0.7756 & 0.6709 & 0.6962 & 0.4503 & 0.4362 & 0.3860 & 0.3813 \\
                TrafficFormer & 0.8912 & 0.7572 & 0.7835 & 0.7530 & 0.7636 & 0.7172 & 0.6411 & 0.6594 & 0.4564 & 0.4296 & 0.3539 & 0.3568 \\
                YaTC & \underline{0.9635} & 0.8578 & \underline{0.8683} & 0.8503 & \textbf{0.8747} & \underline{0.8084} & 0.7661 & 0.7745 & 0.8994 & 0.9241 & 0.8993 & 0.9143 \\
                NetMamba & 0.9626 & \underline{0.8756} & 0.8531 & \underline{0.8638} & 0.8520 & \textbf{0.8122} & \underline{0.7675} & \underline{0.7801} & 0.6725 & 0.8886 & 0.6155 & 0.7018 \\
                \textbf{Traffic-MoE} & \textbf{0.9769} & \textbf{0.8849} & \textbf{0.8860} & \textbf{0.8839} & \underline{0.8588} & 0.8007 & \textbf{0.7701} & \textbf{0.7824} & \textbf{0.9305} & \textbf{0.9321} & \textbf{0.9323} & \textbf{0.9307} \\
                \bottomrule
            \end{tabular}
            
        \end{tabular}
    }
\end{table*}

\subsection{Comparison with State-of-the-Art Methods}
Table~\ref{tab:method_comparison} reports the overall performance of all methods across diverse traffic analysis scenarios, where \model consistently demonstrates highly competitive performance across all evaluation metrics.
A cross-paradigm comparison reveals a substantial performance gap: statistical ML methods and DL models trained from scratch lag considerably behind pre-trained methods on most tasks.
This disparity highlights a fundamental challenge in modeling heterogeneous network traffic. Traditional methods relying on shallow statistical features struggle to capture complex protocol behaviors obscured by encryption. Similarly, training deep networks from scratch fails to infer intricate protocol state transitions and temporal burst patterns given the scarcity of labeled attack samples. In contrast, by leveraging self-supervised pre-training on a massive unlabeled traffic corpus, \model internalizes the intrinsic ``grammar'' of network protocols (\eg handshake sequences and payload structures), providing a semantic-rich initialization that significantly reduces dependency on task-specific labeled traffic data.

In obfuscated Tor encrypted scenarios, \model demonstrates exceptional robustness. As shown in the \textit{ISCXTor2016} results, \model achieves Macro-F1 scores of 0.8900 (NonTor) and 0.8072 (Tor), outperforming the strongest baseline by 4.68\% and 11.28\%, respectively. Notably, existing strong baselines such as ET-BERT and NetMamba suffer from noticeable performance degradation in Tor traffic. This decline stems from Tor protocol's multi-layer encryption mechanisms, which introduce significant structural noise, thereby confusing dense architectures that process all tokens uniformly. \model overcomes this challenge through its sparse MoE architecture. The dynamic routing mechanism allows specific experts to specialize in decoupling valid behavioral fingerprints from padding noise, thereby maintaining high discriminative power even under extreme anonymity constraints.

Similarly, in VPN-encapsulated environments, \model effectively penetrates protocol obfuscation to identify underlying services. On the \textit{ISCXVPN2016} dataset, particularly in the ``Mixed Traffic'' detection task, \model outperforms the generative pre-trained model NetGPT by 10.12\% in accuracy and 3.71\% in Macro-F1. While traditional methods like FlowPrint struggle due to the homogenization of observable statistical features by VPN tunneling, \model succeeds by modeling long-range contextual dependencies. Through the self-attention mechanism, it captures intrinsic flow evolution patterns that remain invariant across tunnel encapsulation, thus resisting the interference of outer protocol headers.

Regarding attack detection, \model exhibits critical advantages in identifying malicious activities within massive background traffic. On the \textit{CICIoMT2024} dataset, \model achieves an accuracy of 0.9769 and outperforms the second-best model by 2.33\% in Macro-F1. 
On the more complex \textit{CICIoT2023} dataset, which comprises 34 categories, \model establishes leading Macro-F1 (0.7824) and Macro-Recall (0.7701). Although Accuracy and Macro-Precision marginally trail YaTC and NetMamba, \model strikes a superior precision-recall trade-off. This indicates that our MoE-driven design effectively mitigates the bias toward majority classes, enabling specialized experts to capture fine-grained nuances of minority attack variants in the long-tail distribution. Consequently, \model yields a more comprehensive detection capability across diverse threat categories.
Furthermore, we observe that traditional methods fail to maintain performance in this scenario due to the severe class imbalance and the presence of diverse attack variants. Pre-trained models address this by internalizing anomalous patterns during the pre-training phase, enabling them to generalize from common attacks to unseen variants and effectively mitigate the long-tail distribution issue inherent in real-world threat detection.

Notably, \model demonstrates superior resilience on \textit{CipherSpectrum}, a challenging pure TLS 1.3 encrypted traffic dataset. While existing pre-trained models, such as ET-BERT, NetGPT, and TrafficFormer, experience severe feature collapse on this dataset (with Macro-F1 plummeting below 0.39), \model establishes a new SOTA, achieving an accuracy of 0.9305 and a Macro-F1 of 0.9307. This performance surpasses both generic pre-trained baselines and statistical fingerprinting methods, validating \model's capacity to effectively overcome representation bottlenecks introduced by modern, highly encrypted traffic environments.

\begin{table*}[!t]
    \centering
    \caption{Ablation study results under different traffic detection tasks}
    \label{tab:ablation_study}
    
    \resizebox{\textwidth}{!}{%
        \begin{tabular}{c cccc cccc cccc cccc}
            \toprule
            \multirow{2}{*}{\textbf{Ablation Variant}} & 
            \multicolumn{4}{c}{\textbf{Task 1: ISCXTor2016(NonTor)}} & 
            \multicolumn{4}{c}{\textbf{Task 2: ISCXVPN2016(Mixed)}} & 
            \multicolumn{4}{c}{\textbf{Task 3: CICIoMT2024}} & 
            \multicolumn{4}{c}{\textbf{Task 4: CipherSpectrum}} \\
            \cmidrule(lr){2-5} \cmidrule(lr){6-9} \cmidrule(lr){10-13} \cmidrule(lr){14-17}
             & ACC & M-PR & M-RC & M-F1 & ACC & M-PR & M-RC & M-F1 & ACC & M-PR & M-RC & M-F1 & ACC & M-PR & M-RC & M-F1 \\
            \midrule
            
            \textbf{Traffic-MoE(Ours)} & \underline{0.9827} & \underline{0.9222} & \textbf{0.8707} & \underline{0.8900} & \textbf{0.7679} & \textbf{0.8306} & \textbf{0.8377} & \textbf{0.8332} & \underline{0.9769} & \underline{0.8849} & \underline{0.8860} & \underline{0.8839} & \textbf{0.9305} & \underline{0.9321} & \textbf{0.9323} & \textbf{0.9307} \\
            MoE$\rightarrow$Dense(Total) & \textbf{0.9831} & \textbf{0.9229} & 0.8655 & \textbf{0.8904} & \underline{0.7600} & \underline{0.8101} & \underline{0.8302} & \underline{0.8193} & \textbf{0.9790} & \textbf{0.8957} & \textbf{0.8947} & \textbf{0.8946} & \underline{0.9300} & \textbf{0.9346} & \underline{0.9233} & \underline{0.9280} \\
            MoE$\rightarrow$Dense(Active)& 0.9806 & 0.9069 & 0.8540 & 0.8758 & 0.7372 & 0.7959 & 0.8011 & 0.7976 & 0.9760 & 0.8783 & 0.8787 & 0.8774 & 0.9249 & 0.9284 & 0.9168 & 0.9222 \\
            w/o Shared Expert & 0.9725 & 0.8814 & 0.7854 & 0.8182 & 0.6636 & 0.7088 & 0.7220 & 0.7143 & 0.9110 & 0.8332 & 0.8344 & 0.8221 & 0.7149 & 0.7438 & 0.6984 & 0.7116 \\
            Dense(Total) w/o PT & 0.9674 & 0.8064 & 0.8054 & 0.7938 & 0.6111 & 0.6958 & 0.6680 & 0.6770 & 0.9728 & 0.8350 & 0.8346 & 0.8294 & 0.8577 & 0.8694 & 0.8499 & 0.8584 \\
            w/o PT & 0.9643 & 0.7753 & 0.7271 & 0.7481 & 0.6255 & 0.6883 & 0.6865 & 0.6854 & 0.9685 & 0.8110 & 0.8236 & 0.8154 & 0.6833 & 0.6842 & 0.6601 & 0.6635 \\
            w/o $\mathcal{L}_\text{aux}$ & 0.9800 & 0.8786 & \underline{0.8658} & 0.8718 & 0.7353 & 0.8064 & 0.8205 & 0.8107 & 0.9767 & 0.8674 & 0.8725 & 0.8692 & 0.9240 & 0.9222 & 0.9162 & 0.9185 \\
            w/o Traffic2Token & 0.9716 & 0.9009 & 0.8403 & 0.8719 & 0.6715 & 0.7663 & 0.7844 & 0.7732 & 0.9722 & 0.8249 & 0.8216 & 0.8109 & 0.9126 & 0.9088 & 0.9074 & 0.9079 \\
            
            \midrule
            \midrule 
            
            \textbf{Traffic-MoE(Ours)} & \textbf{0.9827} & \textbf{0.9222} & \textbf{0.8707} & \textbf{0.8900} & \textbf{0.7679} & \underline{0.8306} & \textbf{0.8377} & \underline{0.8332} & \underline{0.9769} & \textbf{0.8849} & \underline{0.8860} & \textbf{0.8839} & \textbf{0.9305} & \underline{0.9321} & \underline{0.9323} & \textbf{0.9307} \\
            w/ Header Only & 0.9750 & 0.8776 & 0.7983 & 0.8290 & 0.7425 & 0.8050 & 0.8150 & 0.8089 & 0.9263 & 0.8444 & 0.8539 & 0.8481 & 0.9271 & 0.9224 & 0.9253 & 0.9226 \\
            w/ Payload Only & 0.8317 & 0.7535 & 0.7209 & 0.7455 & 0.5199 & 0.7147 & 0.5382 & 0.5648 & 0.4879 & 0.4603 & 0.4459 & 0.4210 & 0.5390 & 0.5175 & 0.4991 & 0.4897 \\
            w/o Payload-Data & \underline{0.9825} & \underline{0.9097} & \underline{0.8684} & \underline{0.8884} & \underline{0.7653} & \textbf{0.8345} & \underline{0.8336} & \textbf{0.8401} & \textbf{0.9813} & \underline{0.8831} & \textbf{0.8912} & \underline{0.8797} & \underline{0.9296} & \textbf{0.9323} & \textbf{0.9343} & \underline{0.9301} \\
            w/o Payload-Negotiation & 0.9096 & 0.8077 & 0.7778 & 0.8064 & 0.6822 & 0.7437 & 0.6709 & 0.6916 & 0.8273 & 0.7536 & 0.7471 & 0.7483 & 0.5893 & 0.6964 & 0.7190 & 0.6449 \\
            w/o Header-Length & 0.9544 & 0.8753 & 0.7741 & 0.7866 & 0.6434 & 0.7140 & 0.6568 & 0.6727 & 0.8914 & 0.6968 & 0.7247 & 0.6794 & 0.5382 & 0.5724 & 0.5027 & 0.4991 \\
            w/o Header-Direction & 0.9750 & 0.8679 & 0.8290 & 0.8412 & 0.7387 & 0.8115 & 0.7862 & 0.7949 & 0.9620 & 0.7498 & 0.7378 & 0.7434 & 0.9225 & 0.9278 & 0.9172 & 0.9208 \\
            w/o Header-Flags & 0.9690 & 0.8445 & 0.8559 & 0.8450 & 0.7227 & 0.7919 & 0.7472 & 0.7668 & 0.7848 & 0.7119 & 0.6579 & 0.6443 & 0.9106 & 0.9205 & 0.9019 & 0.9047 \\
            w/o Header-Proto & 0.9808 & 0.8966 & 0.8580 & 0.8741 & 0.7638 & 0.8305 & 0.8297 & 0.8309 & 0.9701 & 0.7855 & 0.7702 & 0.7773 & 0.9245 & 0.9308 & 0.9284 & 0.9268 \\
            w/o Header-Time & 0.9366 & 0.7722 & 0.7892 & 0.7658 & 0.5632 & 0.7754 & 0.6859 & 0.6599 & 0.7426 & 0.5617 & 0.5500 & 0.5264 & 0.5676 & 0.8503 & 0.5681 & 0.5825 \\
            
            \bottomrule
        \end{tabular}
    }
\end{table*}

\subsection{Comparison in Few-Shot Settings}
\begin{figure}[!t]
    \centering
    \includegraphics[width=\linewidth]{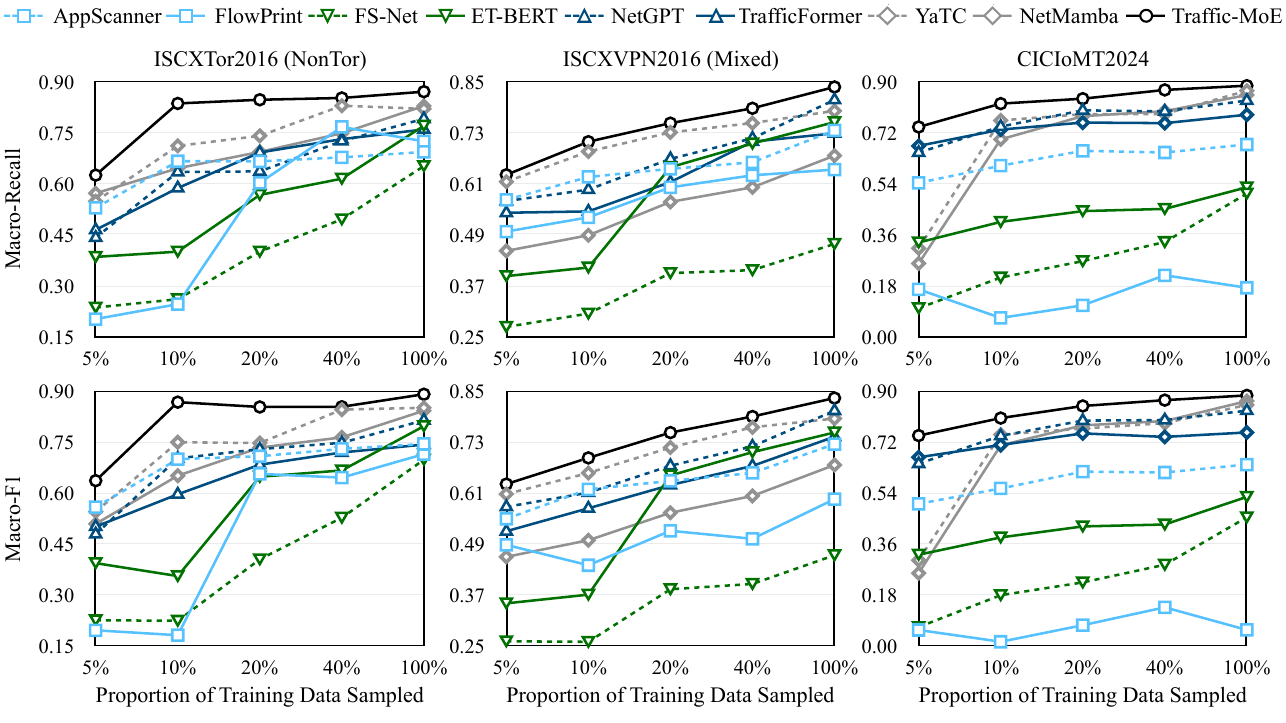}
    \caption{Performance evaluation in few-shot supervision.}
    \label{fig:fewshot}
\end{figure}

In real-world cybersecurity operations, acquiring large-scale, high-quality labeled traffic is often cost-prohibitive due to the rapid evolution of attack variants. To evaluate robustness under \textit{Label Scarcity}, we conduct a stratified few-shot evaluation on three representative datasets (\textit{ISCXTor2016-NonTor}, \textit{ISCXVPN2016-Mixed}, \textit{CICIoMT2024}), scaling training data from 5\% to 100\%. The performance trajectories in Figure~\ref{fig:fewshot} reveal three distinct evolution paradigms. 
Scratch-trained DL methods (\eg FS-Net) exhibit data-hungry collapse, where the lack of prior knowledge leads to parameter non-convergence and precipitous performance drops (over 85\% decline) in extreme few-shot settings. 
In contrast, statistical ML methods (\eg AppScanner) demonstrate stability but suffer from a heuristic representation ceiling, failing to scale performance with increased data due to the limited expressiveness of handcrafted features. 
Most notably, \model achieves \textbf{sample-efficient transfer}, significantly outperforming baselines at the 5\% data extreme and often matching the full-data performance of competitors with only 10\%-20\% supervision, validating that the pre-trained MoE architecture successfully internalizes transfer-ready universal traffic semantics.

A deeper inspection of critical crossovers in Figure~\ref{fig:fewshot} highlights the architectural advantage of sparse experts over dense transformers. In complex encrypted scenarios (\textit{ISCXVPN2016}), we observe a watershed at the 20\% data ratio: below this threshold, dense pre-trained baselines (\eg ET-BERT) suffer significant degradation, dropping below even traditional ML methods. This indicates that dense architectures are prone to \textit{overfitting and representation collapse} when fine-tuning data is insufficient to constrain their massive parameter space. Conversely, \model maintains highly competitive performance, suggesting that \textit{sparse expert activation mechanism effectively acts as a regularizer}. By activating only a subset of parameters ($k=2$) per token, it reduces effective model complexity for specific tasks to prevent overfitting. Furthermore, in intrusion detection tasks (\textit{CICIoMT2024}), \model maintains robust performance even when traditional methods collapse, demonstrating significant \textbf{generalization} potential for identifying emerging threats by effectively transferring learned protocol semantics to attack signatures.

\subsection{Ablation Analysis}
To quantify the contributions of \model's key architectural designs and input flow sequence features, we conduct a multi-dimensional ablation study (results for four representative tasks are summarized in Table~\ref{tab:ablation_study}, with more dataset evaluations detailed in Appendix F).

\subsubsection{Effectiveness of Key Model Components} We construct model variants by replacing or removing critical modules to validate the architectural rationality of \model.
\ding{182} \textbf{Sparse Routing and Expert Specialization} Compared to parameter-matched dense baselines, \model achieves superior performance while maintaining extremely low inference overhead. Specifically, compared to the active-parameter-matched variant ``MoE$\rightarrow$Dense(Active)'', it yields an average Macro-F1 improvement of 1.94\%. In the complex \textit{ISCXVPN2016(Mixed)} scenario, it even outperforms the total-parameter-matched variant ``MoE$\rightarrow$Dense(Total)'' by a margin of 1.70\%. This substantiates that dynamic routing effectively decouples heterogeneous traffic patterns, successfully breaking through the representation bottlenecks inherent in dense architectures.
Furthermore, eliminating the persistently activated shared expert (w/o Shared Expert) induces a consistent performance degradation, particularly in encrypted traffic tasks. This confirms its critical role as an anchor for universal protocol semantics, preventing the representational capacity of specialized experts from being diluted.
\ding{183} \textbf{Necessity of the Pre-training Paradigm:} Removing the pre-training phase (w/o PT) causes an average Macro-F1 plummet of 17.54\% across all tasks. Notably, without pre-training, the performance degradation of the MoE architecture far exceeds that of dense models (\eg dropping by 28.71\% vs. 7.50\% in TLS encrypted scenarios). This stark contrast reveals a critical mechanism: the routing module heavily relies on a robust, pre-established semantic space. Only upon this foundation can the router avoid blind allocation and converge to an optimal state of expert specialization.
\ding{184} \textbf{Load Balancing and Data Processing Pipeline:} Removing the load-balancing loss $\mathcal{L}_\text{aux}$ results in a general Macro-F1 decline of 1.31\% to 2.70\%, indicating that $\mathcal{L}_\text{aux}$ is essential for ensuring balanced expert activation and maximizing the utilization of overall model capacity. Furthermore, compared to directly feeding raw traffic bytes, employing the Traffic2Token processing pipeline yields an average Macro-F1 improvement of 5.34\%, proving that this design effectively mitigates the learning difficulty associated with long sequence features.

\subsubsection{Effectiveness of Traffic Sequence Components} To explore the contributions of multimodal features, we perform fine-grained component masking and disentanglement on the input flow sequences:
\ding{182} \textbf{Modality Complementarity and Phase Awareness:} Macroscopically, header metadata dominates in most tasks (i.e., \textit{w/ Header Only} significantly outperforms \textit{w/ Payload Only}), while \model successfully compensates for single-modality deficiencies through multi-view fusion. Deeper ``payload phase awareness" experiments reveal that masking the \textit{data transfer phase} of the payload (\textit{w/o Payload-Data}) has a negligible impact on performance; however, masking the \textit{handshake negotiation phase} (\textit{w/o Payload-Negotiation}) triggers a sharp Macro-F1 drop of 9.39\% to 30.71\%. This provides conclusive evidence: the performance gain derived from payload features essentially originates from the plaintext interaction structures during handshakes, rather than the semantically opaque encrypted application data.
\ding{183} \textbf{Metadata Feature Importance Quantification:} Quantitative results from masking individual header metadata attributes (using [PAD]) demonstrate that the packet inter-arrival time contributes the most (an average Macro-F1 drop of 28.15\% when removed), whereas the protocol type (Proto) has the minimal impact. Crucially, in highly obfuscated modern encryption scenarios, the discriminative power of length features is significantly amplified; removing these features leads to a catastrophic Macro-F1 degradation of 46.37\%.

\subsection{Inference Efficiency Analysis}
\begin{figure}[!htp]
    \centering
    \includegraphics[width=\linewidth]{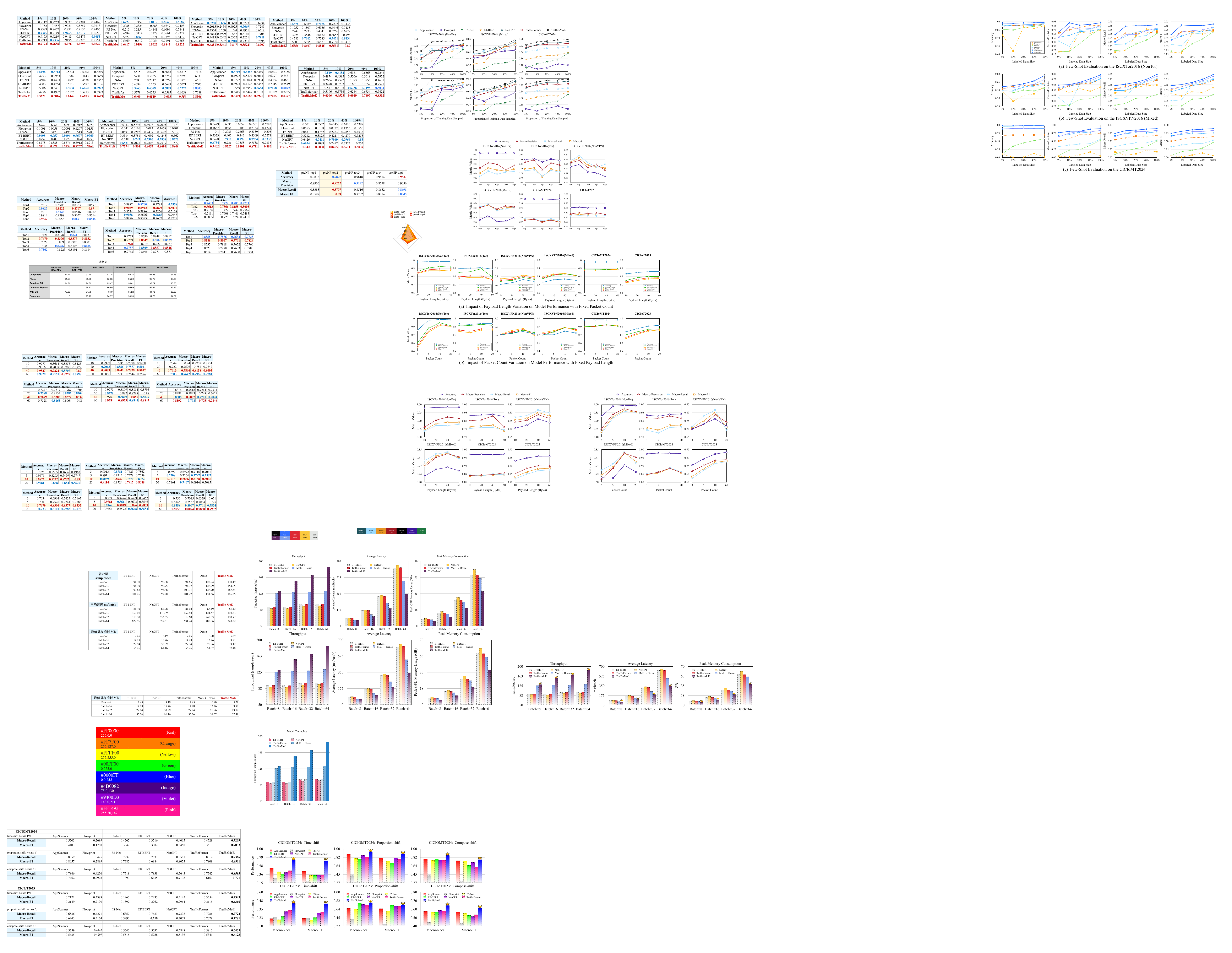}
    \caption{Inference efficiency comparison on CICIoMT2024.}
    \label{fig:inference-effiency}
\end{figure}

To rigorously assess efficiency, we conduct a comprehensive inference efficiency analysis of \model against the total-parameter-matched dense variant (MoE$\rightarrow$Dense) and other competitors on \textit{CICIoMT2024} across varying concurrency loads (batch size $B \in \{8, 16, 32, 64\}$). As illustrated in Figure~\ref{fig:inference-effiency}, \model exhibits a pronounced efficiency advantage that scales non-linearly with traffic load. Specifically, under high-load conditions ($B=64$), \model achieves a 41.57\% improvement in throughput and a 29.36\% reduction in latency compared to the dense variant. This gain stems from decoupling model capacity from inference FLOPs. By dynamically routing tokens to only the top-$2$ experts, \model reduces computational complexity from $\mathcal{O}(N_\text{total})$ to $\mathcal{O}(N_\text{active})$, effectively bypassing approximately 40\% of redundant computation.
Furthermore, while the runner-up NetMamba inherently exhibits lower peak memory consumption due to underlying operator optimizations, \model leverages \textit{activation sparsity} to achieve a vastly superior latency-throughput trade-off, delivering a 70.42\% increase in throughput and a 41.39\% decrease in average latency. This demonstrates that our MoE-driven design effectively circumvents the traditional computational bottlenecks of dense architectures, making it exceptionally well-suited for deployment on high-speed, throughput-sensitive network security gateways.

\begin{figure}[!t]
    \centering
    \includegraphics[width=\linewidth]{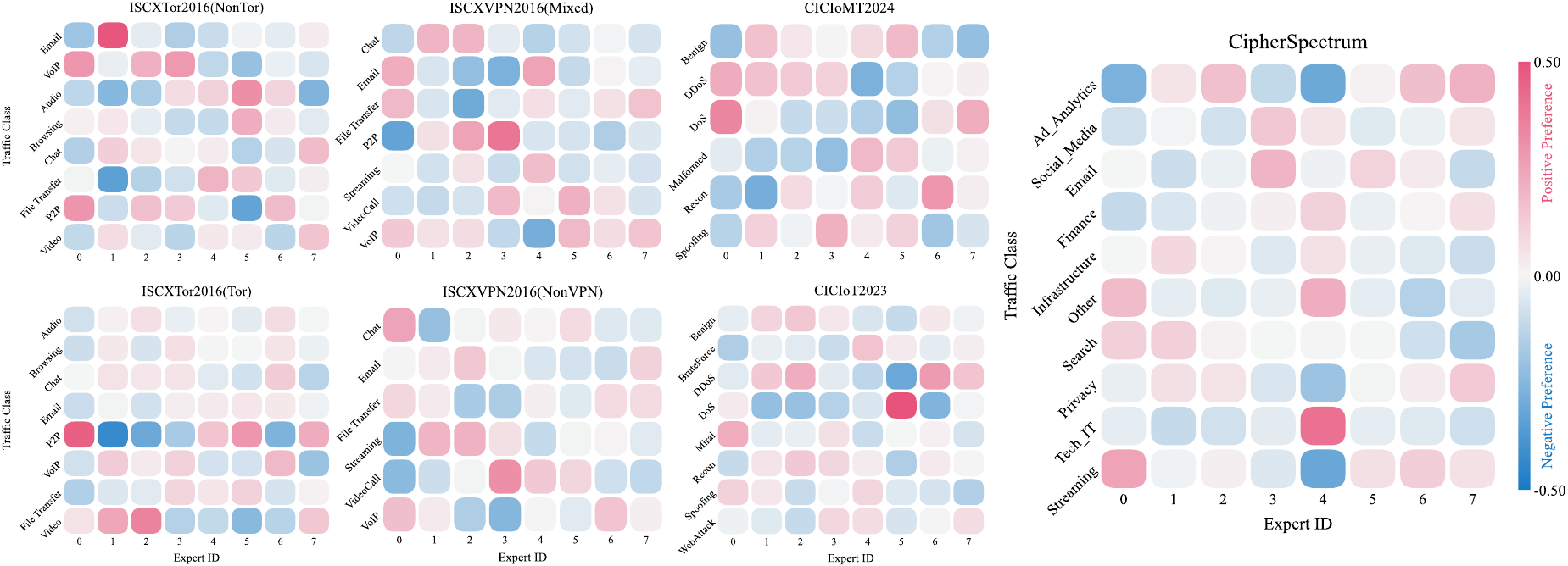}
    \caption{Expert activation preference across detection tasks.}
    \label{fig:expertprefer}
\end{figure}

\subsection{Visual Analysis of Expert Activation Preferences}
To qualitatively reveal the domain specialization of individual experts for specific traffic patterns within the MoE architecture, we perform a visual analysis of the traffic category-to-expert mapping in the deeper layers of the model (as illustrated in Figure~\ref{fig:expertprefer}). Since shallow layers tend to extract generic protocol grammars, our analysis focuses on the deep layers where task-specific specialization occurs.

For quantitative assessment, instead of using raw Softmax probabilities, we compute the empirical activation frequency of each expert across the dataset. To eliminate distribution bias from the global routing baseline, we further calculate the Relative Preference Rate, defined as the relative deviation of category-specific activation frequency from the global average. As indicated by the color scale, warm tones and cool tones quantify an expert's positive preference (over-activation) or active repulsion toward a specific traffic category, respectively.

As shown in Figure~\ref{fig:expertprefer}, the fine-tuned model exhibits highly differentiated routing patterns across four distinct traffic scenarios. The unique combinations of preferred and repelled experts for different categories substantiate that the routing network has learned to dynamically and precisely invoke specific expert subsets based on the intrinsic semantics of the input traffic. Furthermore, while minor expert overlap exists between categories within homogeneous datasets, this phenomenon reflects the model’s strategic parameter reuse and collaborative feature extraction within a constrained parameter space.
Further analysis reveals that a single traffic category typically triggers positive preferences across multiple experts. This collaborative mechanism is facilitated by the auxiliary load-balancing loss $\mathcal{L}_\text{aux}$, which prevents expert saturation and encourages diverse experts to jointly model complex traffic representations. Consequently, the model effectively avoids ``expert collapse" and leverages the sparse parameter space for multi-faceted feature decomposition.

Overall, the visual evidence confirms that experts within MoE layers autonomously evolve into domain experts specialized in downstream task features. By capturing high-level behavioral semantics (\eg encryption entropy distributions, traffic burst patterns), these experts significantly expand the model's decision boundary in complex network environments while maintaining robust generalization.

\subsection{Evaluation under Distribution Shifts}
\begin{figure}[!t]
    \centering
    \includegraphics[width=\linewidth]{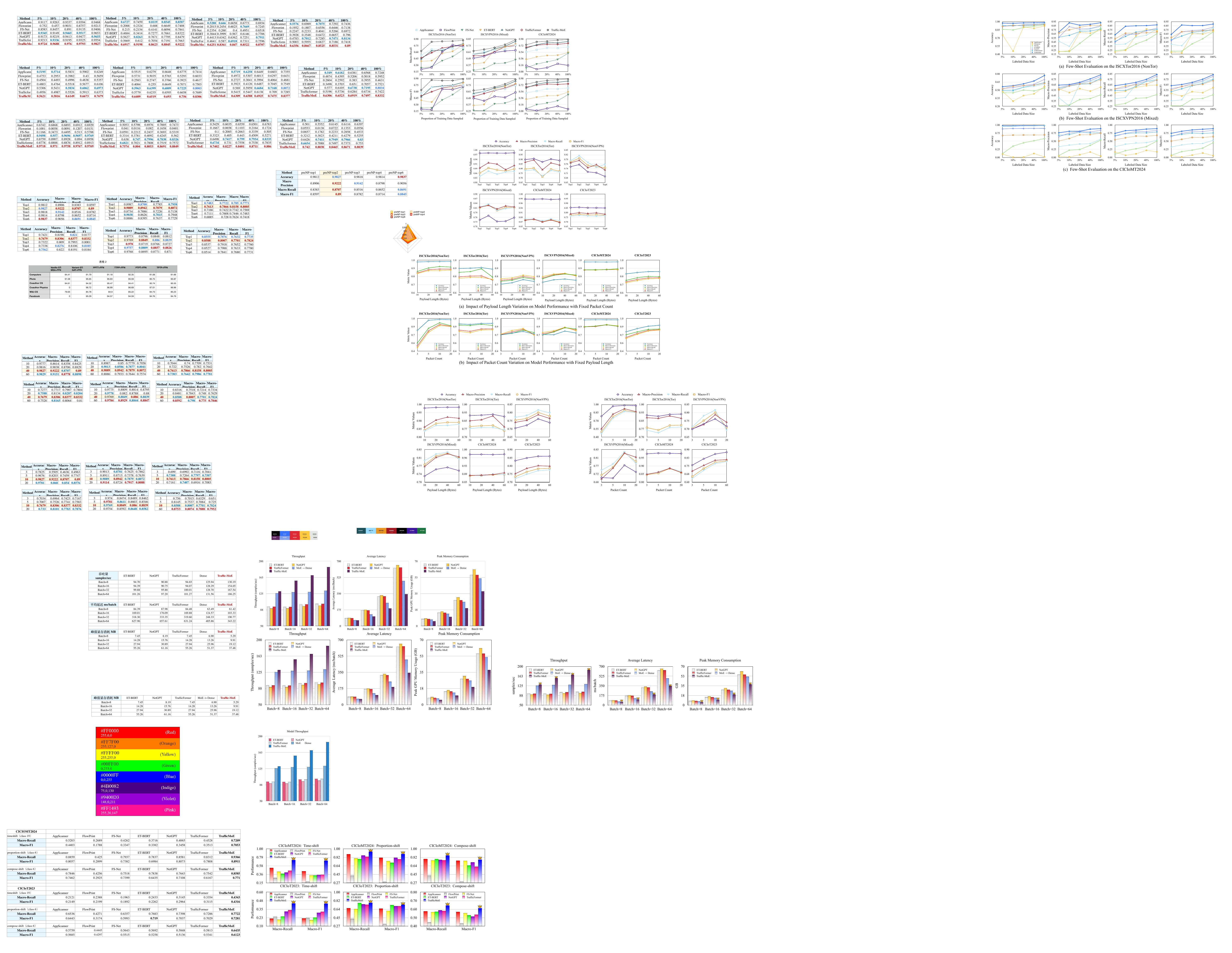}
    \caption{Performance under different distribution shifts.}
    \label{fig:ood}
\end{figure}

Real-world network environments are highly dynamic, characterized by continuous temporal evolution and non-stationary class distributions. To evaluate whether \model can adapt to such ``open-world'' challenges without frequent retraining, we construct three OOD benchmarks: \textbf{Time-shift} (simulating concept drift over time), \textbf{Proportion-shift} (simulating intra-class distribution shift), and \textbf{Compose-shift} (simulating unseen attack sub-variants). Please refer to Appendix I for more details.
Figure~\ref{fig:ood} visualizes the performance comparison across these scenarios.
\ding{182} \textbf{Resilience to Concept Drift:} The most significant performance gap appears in the Time-shift scenario. While existing competitors (\eg FS-Net, ET-BERT) suffer catastrophic degradation due to temporal feature shifts, \model maintains a robust performance, outperforming the runner-up by 32.31\%--60.18\% across datasets. This suggests that the sparse expert architecture effectively disentangles time-invariant protocol semantics from transient statistical noise, preventing the model from overfitting to temporal shortcuts. 
\ding{183} \textbf{Stability under Structural Shifts:} In scenarios involving structural mutations (Proportion/Compose-shift), \model consistently achieves competitive performance. Notably, in the Compose-shift task (where 50\% of sub-classes are unseen during training), \model surpasses strong pre-trained competitors (\eg NetGPT, YaTC) by significant margins. High recall rates in these tasks indicate that the model learns generalized behavioral abstractions rather than memorizing specific signatures of seen sub-classes.
In summary, \model exhibits superior generalization and robustness against distribution shifts, validating its viability for sustained operation in evolving network defense systems.

\section{Conclusion}
This work bridges the gap between advanced representation learning and practical network protection by introducing \model, an efficient sparse foundation model. By synergizing self-supervised pre-training with a dynamic MoE architecture, \model effectively decouples model capacity from inference overhead. Extensive evaluations demonstrate that \model establishes highly competitive performance across heterogeneous security tasks including intrusion detection and encrypted service classification while significantly reducing inference latency compared to dense competitors. Furthermore, its superior robustness and generalization under few-shot supervision and distribution shifts confirms that sparse foundation models offer a scalable, resilient, and efficient paradigm for next-generation real-time network defense. 

\bibliographystyle{IEEEtran} 
\bibliography{mybib,IEEEabrv}

\vspace{-60pt}
\begin{IEEEbiography}[{\includegraphics[width=1in,height=1.25in,clip,keepaspectratio]{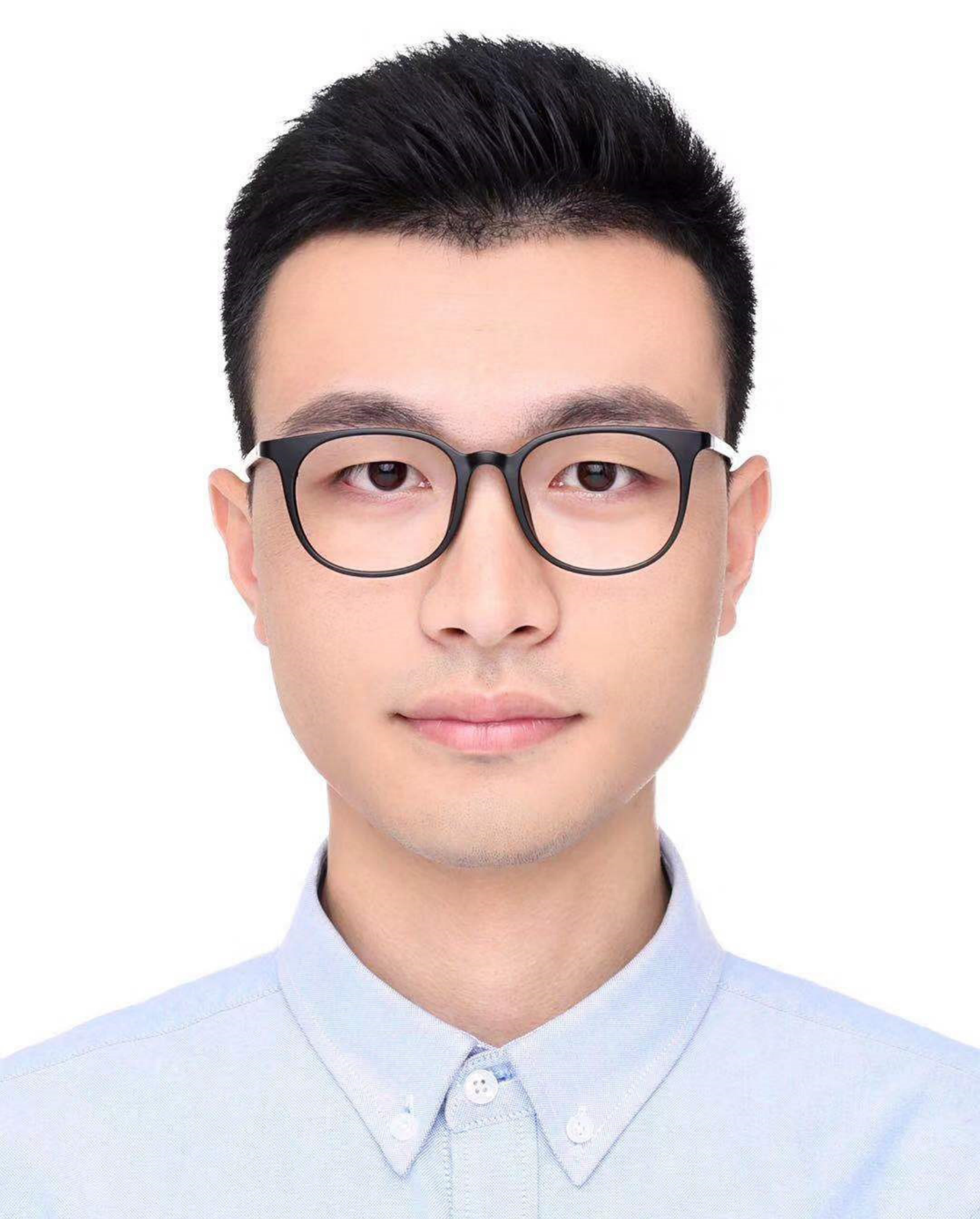}}]{Jiajun Zhou}
    received the Ph.D degree in control theory and engineering from Zhejiang University of Technology, Hangzhou, China, in 2023. He is currently an Associate Research Fellow with the Institute of Cyberspace Security, Zhejiang University of Technology. His current research interests include graph data mining, cyberspace security and AI security.
\end{IEEEbiography}
\vspace{-20pt}

\begin{IEEEbiography}[{\includegraphics[width=1in,height=1.25in,clip,keepaspectratio]{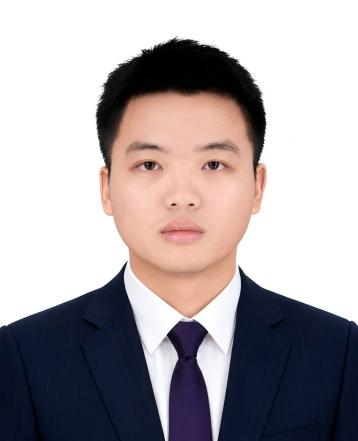}}]{Changhui Sun}
    received the B.S. degree in network engineering from Wenzhou University, Wenzhou, China, in 2024. He is currently pursuing the M.S. degree in computer science and technology at Zhejiang University of Technology, Hangzhou, China. His current research interests include cyberspace security and network traffic detection.
\end{IEEEbiography}
\vspace{-20pt}

\begin{IEEEbiography}[{\includegraphics[width=1in,height=1.25in,clip,keepaspectratio]{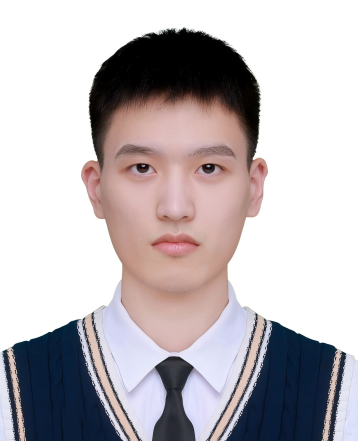}}]{Wentao Fu}
    received the B.S. degree in Automation from Xi'an University of Posts and Telecommunications, Xi'an, China, in 2023. He is currently pursuing his master's degree at the Institute of Cyberspace Security, Zhejiang University of Technology, China. His current research interests include cyberspace security and network intrusion detection.
\end{IEEEbiography}
\vspace{-20pt}

\begin{IEEEbiography}[{\includegraphics[width=1in,height=1.25in,clip,keepaspectratio]{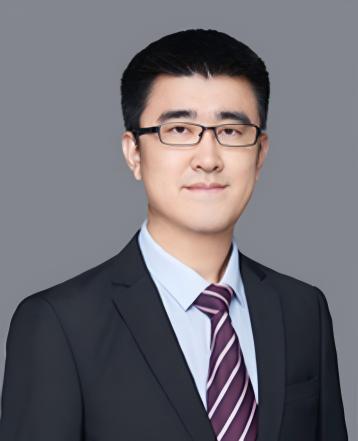}}]{Meng Shen}
    (Member, IEEE) received the B.Eng. degree in computer science from Shandong University, Jinan, China, in 2009, and the Ph.D. degree in computer science from Tsinghua University, Beijing, China, in 2014. He is currently a Professor at Beijing Institute of Technology, Beijing. He has authored more than 50 papers in top-level journals and conferences, such as USENIX Security, ACM CCS, IEEE Journal on Selected Areas in Communications, and IEEE Transactions on Information Forensics and Security. His research interests include data privacy and security, blockchain applications, and encrypted traffic classification. He received the Best Paper Runner-Up Award from IEEE IPCCC 2014 and IEEE/ACM IWQoS 2020. He was selected by the Beijing Nova Program 2020 and the winner of the ACM SIGCOMM China Rising Star Award 2019. He has guest edited Special Issues on Emerging Technologies for Data Security and Privacy in IEEE Network and IEEE Internet of Things Journal.
\end{IEEEbiography}
\vspace{-20pt}

\begin{IEEEbiography}[{\includegraphics[width=1in,height=1.25in,clip,keepaspectratio]{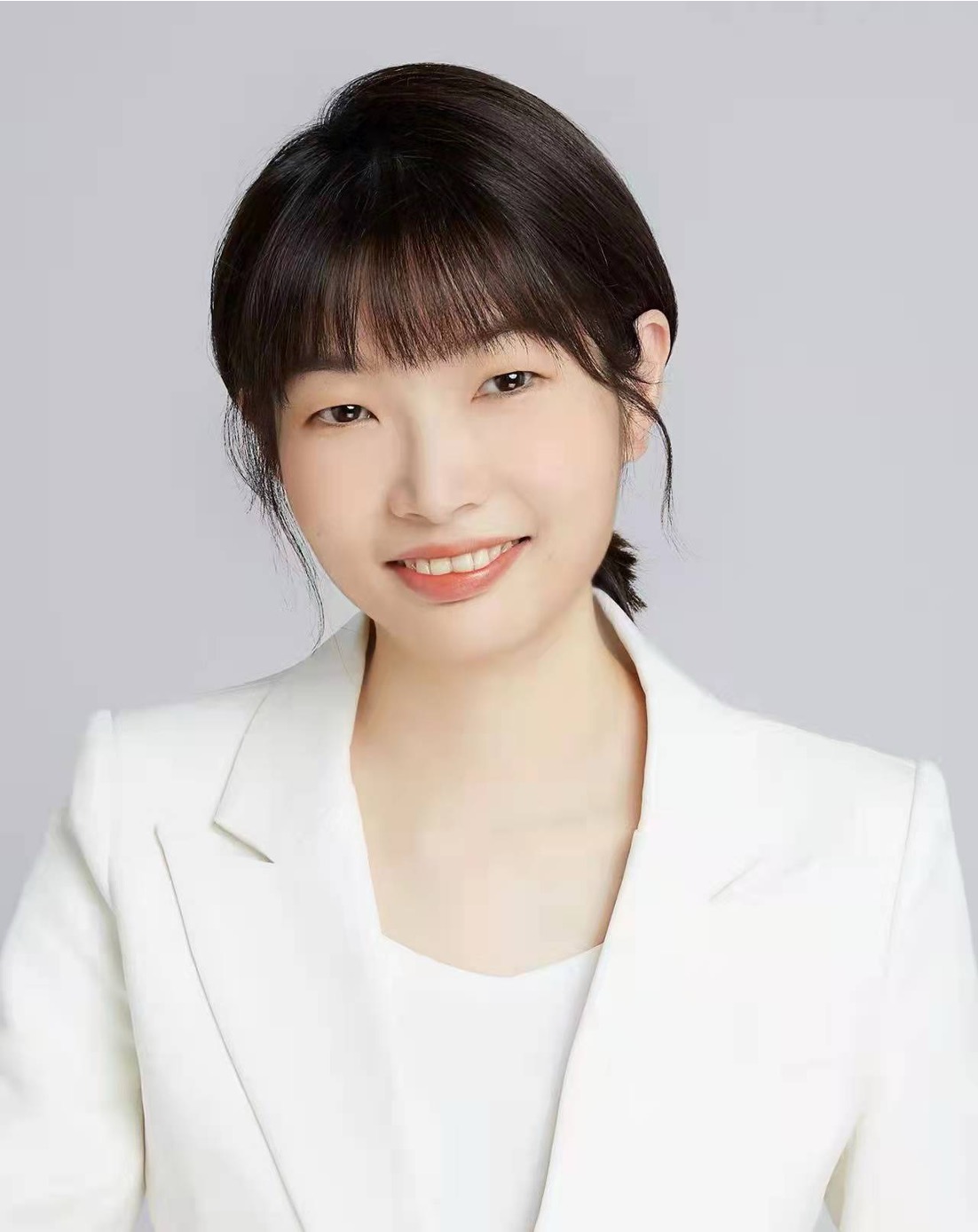}}]{Shanqing Yu}
	received the M.S. degree from the School of Computer Engineering and Science, Shanghai University, China, in 2008 and received the M.S. degree from the Graduate School of Information, Production and Systems, Waseda University, Japan, in 2008, and the Ph.D. degree, in 2011, respectively. She is currently a Lecturer at the Institute of Cyberspace Security and the College of Information Engineering, Zhejiang University of Technology, Hangzhou, China. Her research interests cover intelligent computation and data mining.
\end{IEEEbiography}
\vspace{-20pt}

\begin{IEEEbiography}[{\includegraphics[width=1in,height=1.25in,clip,keepaspectratio]{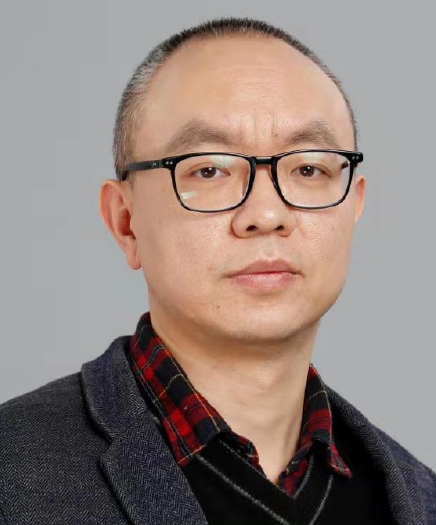}}]{Qi Xuan}(M'18) received the BS and PhD degrees in control theory and engineering from Zhejiang University, Hangzhou, China, in 2003 and 2008, respectively. He was a Post-Doctoral Researcher with the Department of Information Science and Electronic Engineering, Zhejiang University, from 2008 to 2010, respectively, and a Research Assistant with the Department of Electronic Engineering, City University of Hong Kong, Hong Kong, in 2010 and 2017. From 2012 to 2014, he was a Post-Doctoral Fellow with the Department of Computer Science, University of California at Davis, CA, USA. He is a senior member of the IEEE and is currently a Professor with the Institute of Cyberspace Security, College of Information Engineering, Zhejiang University of Technology, Hangzhou, China. His current research interests include network science, graph data mining, cyberspace security, machine learning, and computer vision.
\end{IEEEbiography}

\newpage

\input{appendix}

\end{document}

%% file: appendix.tex
\appendix


\subsection{Pseudocode of Model Training}  
Algorithm~\ref{alg: traffic-moe} outlines the unified training pipeline of \model, detailing the progression from self-supervised pre-training to task-specific fine-tuning.

\begin{algorithm}[!htp]
    \caption{~Pretraining and Fine-tuning of \model.}
    \label{alg: traffic-moe}
    \begin{algorithmic}[1]
    \STATE \textbf{Input:} batch of flow token sequences $\{\boldsymbol{X}^{(b)}\}_{b=1}^{B}$, 
    training mode $\bigstar \in \{\textsc{Pretrain},\textsc{Finetune}\}$, labels $\{y_{b}\}_{b=1}^{B}$ if $\textsc{Finetune}$;
    \STATE \textbf{hyperparameters:} transformer depth $L$, number of experts $N$, top-$k$ routing $k$, auxiliary weight $\lambda_{\mathrm{aux}}$;
    \STATE \textbf{Output:} updated model and task-specific head;
     \item[]  \hfill{\gray{\textsf{/* Backbone: causal transformer with sparse MoE */}}}
    \FOR{$b = 1$ to $B$}
        \STATE Initialize hidden states for the input token sequences: $\ \boldsymbol{H}^{(0,b)} \leftarrow \boldsymbol{X}^{(b)}$;
        \FOR{$l = 1$ to $L$}
        \STATE Compute causal self-attention $\boldsymbol{O}_\mathrm{attn}^{(l,b)}$ via Eq.(1)$\sim$(2);
        \STATE Residual update:
                $\ \widetilde{\boldsymbol{H}}^{(l,b)} \leftarrow \boldsymbol{H}^{(l-1,b)} + \boldsymbol{O}_\mathrm{attn}^{(l,b)}$;
        \STATE Apply sparse MoE to compute $\boldsymbol{H}^{(l,b)}$ via Eq.(4)$\sim$(8);
        \ENDFOR
    \ENDFOR
    \STATE Compute MoE load-balancing loss $\mathcal{L}_{\text{aux}}$ via Eq.(9);
    \item[]  \hfill{\gray{\textsf{/* Pretraining */}}}
    \IF{$\bigstar$ is \textsc{Pretrain}}
        \STATE Compute autoregressive loss $\mathcal{L}_{\text{NP}}$ via Eq.(11); 
        \STATE Combine objectives:
                $\mathcal{L}_{\mathrm{pretrain}} \leftarrow \mathcal{L}_{\mathrm{NP}} + \lambda_{\mathrm{aux}}\cdot\mathcal{L}_{\mathrm{aux}}$;
        \STATE Update backbone parameters by minimizing $\mathcal{L}_{\mathrm{pretrain}}$;
    \item[]  \hfill{\gray{\textsf{/* Fine-tuning */}}}
    \ELSE
        \STATE For each flow $\boldsymbol{X}^{(b)}$, perform mean pooling over final-layer tokens and classify to obtain $p_\theta(y_b \mid \boldsymbol{X}^{(b)})$;
        \STATE Compute classification loss $\mathcal{L}_{\text{TC}}$ via Eq.(14);
        \STATE Combine objectives:
                $\mathcal{L}_{\mathrm{finetune}} \leftarrow \mathcal{L}_{\mathrm{TC}} + \lambda_{\mathrm{aux}}\cdot\mathcal{L}_{\mathrm{aux}}$;
        \STATE Update backbone and prediction head by minimizing $ \mathcal{L}_{\mathrm{finetune}}$ with layer-wise learning-rate decay;
    \ENDIF
    \end{algorithmic}
\end{algorithm}

\subsection{Fine-tuning Datasets}\label{app:dataset}
To systematically evaluate the transfer learning capabilities of model across diverse dimensions, we employ five authoritative public datasets during the fine-tuning phase.
\begin{itemize}[leftmargin=*]
    \item \textbf{CICIoT2023}~\cite{CICIoT2023}: Collected from a complex network environment comprising 105 real IoT devices, this dataset contains benign background traffic and 33 attack behaviors. These attacks are grouped into 7 macro-categories, including DDoS, Brute Force, Web attacks, and Mirai botnets.
    \item \textbf{CICIoMT2024}~\cite{CICIoMT2024}: Derived from a testbed of 40 real and simulated IoMT devices, this dataset encompasses 18 targeted attacks (grouped into 5 categories) exploiting standard healthcare protocols such as Wi-Fi and MQTT.
    \item \textbf{ISCXVPN2016}~\cite{ISCXVPN2016}: This dataset comprises both regular and VPN-encapsulated session traffic. It covers 14 fine-grained traffic categories (\eg VoIP, P2P, Email) generated by diverse applications like Facebook, Skype, and Chrome.
    \item \textbf{ISCXTor2016}\cite{ISCXTor2016}: This dataset includes regular and Tor-obfuscated session traffic. It spans 8 service categories involving applications such as YouTube, Firefox, and Skype.
    \item \textbf{CipherSpectrum}\cite{cipherspectrum}: This dataset features 2024-collected TLS 1.3 encrypted traffic originating from 40 web domains. All sessions are strictly encrypted using the three mainstream AEAD cipher suites mandated by TLS 1.3 protocol.
\end{itemize}

\subsection{Details of Comparison Methods}\label{app:baselines}
\begin{itemize}[leftmargin=10pt]
    \item \textbf{AppScanner}~\cite{AppScanner}: A classic statistical fingerprinting method. It extracts 54 statistical features from encrypted traffic and employs a Random Forest classifier to identify applications without payload inspection. 
    \item \textbf{FlowPrint}~\cite{Flowprint}: A semi-supervised approach designed for open-world scenarios. It mines temporal correlations within traffic to construct a dynamic fingerprint library, enabling the identification of previously unseen applications.
    \item \textbf{FS-Net}~\cite{FS-Net}: An end-to-end deep learning architecture tailored for sequence modeling. It utilizes a Bidirectional Gated Recurrent Unit in an encoder-decoder structure to learn latent representations directly from raw packet length sequences, bypassing manual feature engineering.
    \item \textbf{ET-BERT}~\cite{ET-BERT}: A pioneering method adapting the BERT architecture to encrypted traffic domain. It introduces two pre-training objectives (Masked Burst Model and Same-origin Burst Prediction) to learn contextual dependencies from massive unlabeled flow data.
    \item \textbf{NetGPT}~\cite{NetGPT}: The first framework applying Generative Pre-trained Transformers (GPT) to the traffic domain. It treats network flows as hexadecimal byte sequences and unifies traffic understanding and generation tasks, adapting to downstream tasks via prompt tuning.
    \item \textbf{TrafficFormer}~\cite{TrafficFormer}:
    A recent advancement building upon ET-BERT. It introduces a Same-Origin-Direction-Flow multi-task objective to capture fine-grained packet directionality and sequence relationships more effectively.
    \item \textbf{YaTC}~\cite{YaTC}: A Transformer-based traffic classification model utilizing Masked Autoencoder and multi-level flow representations. It employs dual-level attention mechanisms to extract features, achieving efficient encrypted traffic classification via the pre-training paradigm.
    \item \textbf{NetMamba}~\cite{netmamba}: An efficient pre-trained traffic classification model using unidirectional Mamba. Combined with optimized traffic representation schemes, it significantly accelerates inference while maintaining robust accuracy.
\end{itemize}
\textit{Note on Implementation:} To ensure a fair comparison of \textit{model capacity} rather than data engineering tricks, we standardize the experimental setting. We exclude method-specific data augmentation strategies used in the original papers and train all models on identical raw datasets.

\subsection{Fine-grained Categories to Merged Classes}
Table~\ref{apptab:map class} details the mapping from fine-grained categories to merged classes for \textit{CICIoMT2024} and \textit{CICIoT2023}. 

\begin{table}[!htp]
    \centering
    \renewcommand\arraystretch{1.4}
    \caption{Attack category mapping scheme for the CICIoMT2024 and CICIoT2023 Datasets.}
    \label{apptab:map class}
    \resizebox{\linewidth}{!}{%
    \begin{tabular}{lcc} 
    \toprule
    \multirow{10}{*}{\rotatebox{90}{CICIoMT2024}} & \textbf{Merged Categories} & \textbf{Fine-grained Categories}                                                                                                                                                                                                                                                       \\ 
    \cmidrule{2-3}
                                           & Benign                     & BenignTraffic                                                                                                                                                                                                                                                                          \\
                                           \cdashline{2-3}
                                           & DDoS                       & \begin{tabular}[c]{@{}c@{}}MQTT-DDoS-Connect\_Flood, MQTT-DDoS-Publish\_Flood, TCP\_IP-DDoS-ICMP, \\TCP\_IP-DDoS-SYN, TCP\_IP-DDoS-TCP, TCP\_IP-DDoS-UDP\end{tabular}                                                                                                                  \\
                                           \cdashline{2-3}
                                           & DoS                        & \begin{tabular}[c]{@{}c@{}}MQTT-DoS-Connect\_Flood, MQTT-DoS-Publish\_Flood, TCP\_IP-DoS-ICMP, \\TCP\_IP-DoS-SYN, TCP\_IP-DoS-TCP, TCP\_IP-DoS-UDP\end{tabular}                                                                                                                        \\
                                           \cdashline{2-3}
                                           & Malformed                  & MQTT-Malformed\_Data                                                                                                                                                                                                                                                                   \\
                                           \cdashline{2-3}
                                           & Recon                      & Recon-OS\_Scan, Recon-Ping\_Sweep, Recon-Port\_Scan, Recon-VulScan                                                                                                                                                                                                                     \\
                                           \cdashline{2-3}
                                           & Spoofing                   & ARP\_Spoofing                                                                                                                                                                                                                                                                          \\ 
    \midrule
    \midrule
    \multirow{14}{*}{\rotatebox{90}{CICIoT2023}}  & \textbf{Merged Categories} & \textbf{Fine-grained Categories}                                                                                                                                                                                                                                                       \\ 
    \cmidrule{2-3}
                                           & Benign                     & BenignTraffic                                                                                                                                                                                                                                                                          \\
                                           \cdashline{2-3}
                                           & BruteForce                 & DictionaryBruteForce                                                                                                                                                                                                                                                                   \\
                                           \cdashline{2-3}
                                           & DDoS                       & \begin{tabular}[c]{@{}c@{}}DDoS-ACK-Fragmentation, DDoS-HTTP-Flood, DDoS-ICMP-Flood, \\DDoS-ICMP-Fragmentation, DDoS-PSHACK-Flood, DDoS-RSTFINFlood, \\DDoS-SYN-Flood, DDoS-SlowLoris, DDoS-SynonymousIP-Flood, \\DDoS-TCP-Flood, DDoS-UDP-Flood, DDoS-UDP-Fragmentation\end{tabular}  \\
                                           \cdashline{2-3}
                                           & DoS                        & DoS-HTTP-Flood, DoS-SYN-Flood, DoS-TCP-Flood, DoS-UDP-Flood                                                                                                                                                                                                                            \\
                                           \cdashline{2-3}
                                           & Mirai/Botnet               &  Mirai-greeth-flood, Mirai-greip-flood, Mirai-udpplain                                                                                                                                                                                                                 \\
                                           \cdashline{2-3}
                                           & Recon                      & \begin{tabular}[c]{@{}c@{}}Recon-HostDiscovery, Recon-OSScan, Recon-PingSweep, \\Recon-PortScan, VulnerabilityScan\end{tabular}                                                                                                                                                        \\
                                           \cdashline{2-3}
                                           & Spoofing                   & DNS-Spoofing, MITM-ArpSpoofing                                                                                                                                                                                                                                                         \\
                                           \cdashline{2-3}
                                           & WebAttack                  & \begin{tabular}[c]{@{}c@{}}BackdoorMalware, BrowserHijacking, CommandInjection, \\SqlInjection, Uploading-Attack, XSS\end{tabular}                                                                                                                                                                                               \\
    \bottomrule
    \end{tabular}}
    \end{table}

\subsection{Efficacy and Robustness Analysis}
\begin{figure}[!t]
    \centering
    \includegraphics[width=\linewidth]{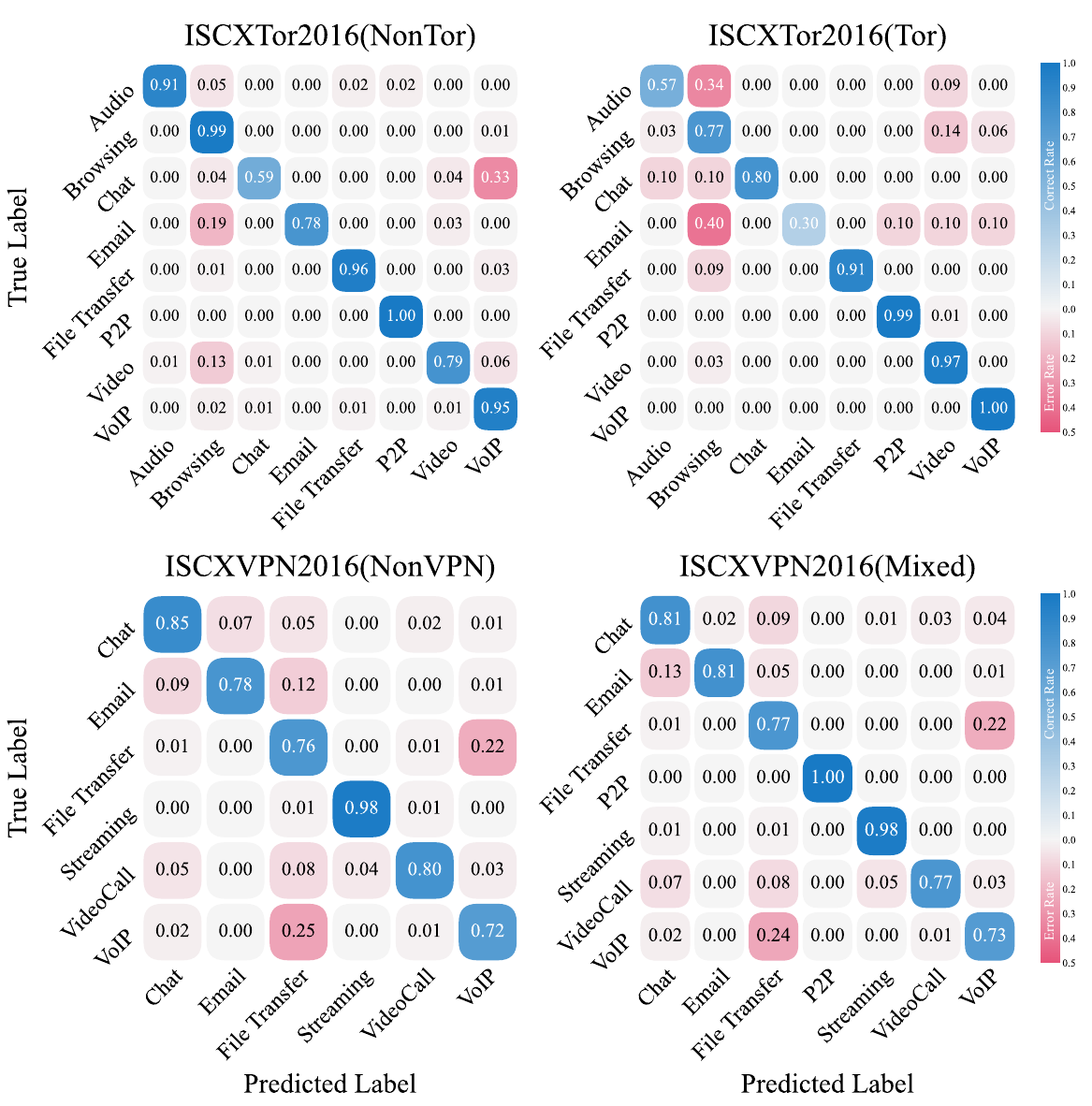}
    \caption{Performance confusion matrices of \model across different traffic classification tasks.}
    \label{fig:confuse}
\end{figure}

To investigate the performance boundaries of \model in real-world security scenarios, we conducts an in-depth analysis of fine-grained results, as shown in Table~\ref{tab:diff-class} and Figure~\ref{fig:confuse}.

\subsubsection{Intrusion Detection: Minimizing Security Risks}
In the context of network defense, the metrics for the \textit{Benign} category serve as global indicators for system reliability: the \textbf{Benign FPR} represents the \textit{Overall Attack Miss Rate} (attacks misclassified as benign), while the \textbf{Benign FNR} represents the \textit{False Alarm Rate} (benign traffic misclassified as attacks). \model demonstrates exceptional capability in preventing defense penetration, achieving a negligible Attack Miss Rate (Benign FPR) of 0.11\% on the sensitive \textit{CICIoMT2024} dataset and a robust 3.14\% on the massive \textit{CICIoT2023} dataset. This implies that the vast majority of threats are successfully intercepted. Conversely, the False Alarm Rate (Benign FNR) is observed at 11.18\% in IoT scenarios. This reflects a ``Fail-Secure'' design philosophy: given the intrinsic behavioral overlap between aggressive benign polling and malicious reconnaissance in IoT networks, the model prioritizes flagging ambiguous traffic for inspection rather than risking a breach. While volumetric attacks are detected with high precision, stealthier categories like \textit{Spoofing} exhibit performance dips, suggesting that low-signature attacks remain a challenge for generalized representations.
\begin{table}
    \centering
    \caption{Detection performance of \model on different traffic categories (merged) in CICIoMT2024 and CICIoT2023}
    \label{tab:diff-class}
    \resizebox{\linewidth}{!}{%
    \begin{tabular}{cccccc} 
    \toprule
    \multirow{7}{*}{\rotatebox{90}{CICIoMT2024~~}} & \textbf{Merged Category} & \textbf{Num.\#Test} & \textbf{Recall} & \textbf{FNR} & \textbf{FPR}  \\ 
    \cmidrule{2-6}
                                           & Benign            & 130                   & 0.9462          & 0.0538       & 0.0011        \\
                                           & DDoS              & 4193                  & 0.9955          & 0.0045       & 0.0046        \\
                                           & DoS               & 4075                  & 0.9931          & 0.0069       & 0.0029        \\
                                           & Malformed         & 45                    & 0.9333          & 0.0667       & 0.0001        \\
                                           & Recon             & 2229                  & 0.9928          & 0.0072       & 0.0013        \\
                                           & Spoofing          & 67                    & 0.7015          & 0.2985       & 0.0019        \\ 
    \midrule
    \midrule
    \multirow{9}{*}{\rotatebox{90}{CICIoT2023~~}}  & \textbf{Merged Category} & \textbf{Num.\#Test} & \textbf{Recall} & \textbf{FNR} & \textbf{FPR}  \\ 
    \cmidrule{2-6}
                                           & Benign            & 3534                  & 0.8882          & 0.1118       & 0.0314        \\
                                           & BruteForce        & 265                   & 0.7925          & 0.2075       & 0.0008        \\
                                           & DDoS              & 10775                 & 0.9426          & 0.0574       & 0.0330        \\
                                           & DoS               & 4000                  & 0.9785          & 0.0215       & 0.0021        \\
                                           & Mirai/Botnet      & 663                   & 0.6591          & 0.3409       & 0.0084        \\
                                           & Recon             & 3754                  & 0.8426          & 0.1574       & 0.0289        \\
                                           & Spoofing          & 2000                  & 0.8250          & 0.1750       & 0.0125        \\
                                           & WebAttack         & 527                   & 0.6831          & 0.3169       & 0.0045        \\
    \bottomrule
    \end{tabular}}
\end{table}

\subsubsection{Tor Anonymity: Countering Traffic Shaping}\label{sec: tor}
In adversarial anonymity scenarios, \model demonstrates a unique ability to counter traffic shaping mechanisms. These operations of fixed-size cell padding and randomized delays used by Tor protocol effectively distort the statistical feature space, leading to significant performance degradation for latency-sensitive services like \textit{Audio} and transactional services like \textit{Email} (as shown in Figure~\ref{fig:confuse}). However, \model exhibits resilience in interactive protocols such as \textit{P2P} and \textit{VoIP}, where detection accuracy remains high. This performance dichotomy indicates that the MoE architecture successfully shifts attention from eroded packet-length statistics to \textit{sequence-level behavioral fingerprints} (\eg handshake orders and interaction frequencies), effectively learning a ``Protocol Grammar'' that persists despite cryptographic obfuscation. 

\subsubsection{VPN Tunneling: Penetrating Encapsulation}
Similarly, under VPN tunneling conditions (\textit{ISCXVPN2016}), \model demonstrates strong resilience to protocol encapsulation. The confusion matrix reveals that misclassifications (\eg between VoIP and File Transfer) are consistent across Non-VPN and Mixed scenarios, stemming from intrinsic feature similarity rather than tunnel noise. Crucially, for complex classes like Streaming and P2P, the model maintains near-perfect identification. This suggests that the self-attention mechanism effectively penetrates the outer tunnel headers to capture long-range flow dependencies, achieving semantic decoupling of the payload behavior from the transport encapsulation.

\subsection{Extended Ablation Analysis}\label{app:ablation}
Table~\ref{apptab:appendix_ablation} provides additional ablation results across three traffic detection tasks.

\begin{table*}[!t]
    \centering
    \caption{More ablation study results across different traffic detection tasks (Appendix)}
    \label{apptab:appendix_ablation}
    
    \resizebox{\textwidth}{!}{%
        \begin{tabular}{c cccc cccc cccc}
            \toprule
            \multirow{2}{*}{\textbf{Ablation Variant}} & 
            \multicolumn{4}{c}{\textbf{Task 1: ISCXTor2016(Tor)}} & 
            \multicolumn{4}{c}{\textbf{Task 2: ISCXVPN2016(NonVPN)}} & 
            \multicolumn{4}{c}{\textbf{Task 3: CICIoT2023}} \\
            \cmidrule(lr){2-5} \cmidrule(lr){6-9} \cmidrule(lr){10-13}
             & ACC & M-PR & M-RC & M-F1 & ACC & M-PR & M-RC & M-F1 & ACC & M-PR & M-RC & M-F1 \\
            \midrule
            
            \textbf{Traffic-MoE(Ours)} & \textbf{0.9089} & \textbf{0.8942} & \textbf{0.7879} & \textbf{0.8072} & \textbf{0.7613} & \textbf{0.7866} & \textbf{0.8158} & \textbf{0.8005} & \underline{0.8588} & \underline{0.8007} & \underline{0.7701} & \underline{0.7824} \\
            MoE$\rightarrow$Dense(Total) & \underline{0.9038} & 0.8656 & 0.7638 & 0.7767 & 0.7178 & 0.7570 & 0.7900 & 0.7713 & \textbf{0.8594} & \textbf{0.8090} & \textbf{0.7765} & \textbf{0.7894} \\
            MoE$\rightarrow$Dense(Active) & 0.8962 & 0.8676 & \underline{0.7774} & \underline{0.7930} & 0.7258 & 0.7632 & 0.7853 & 0.7696 & 0.8521 & 0.7878 & 0.7607 & 0.7704 \\
            w/o Shared Expert & 0.8557 & 0.7433 & 0.6846 & 0.6855 & 0.6268 & 0.6711 & 0.6470 & 0.6440 & 0.7627 & 0.7208 & 0.6798 & 0.6820 \\
            w/o PT & 0.8304 & 0.7071 & 0.6392 & 0.6311 & 0.6062 & 0.6327 & 0.6095 & 0.6153 & 0.7833 & 0.6757 & 0.6584 & 0.6640 \\
            Dense(Total) w/o PT & 0.8506 & 0.7056 & 0.6541 & 0.6564 & 0.6074 & 0.6334 & 0.6533 & 0.6389 & 0.7867 & 0.7054 & 0.6586 & 0.6741 \\
            w/o $\mathcal{L}_\text{aux}$ & 0.8962 & \underline{0.8734} & 0.7683 & 0.7913 & \underline{0.7347} & \underline{0.7803} & \underline{0.7949} & \underline{0.7829} & 0.8535 & 0.8002 & 0.7630 & 0.7809 \\
            w/o Traffic2Token & 0.8835 & 0.8273 & 0.7460 & 0.7645 & 0.6537 & 0.7259 & 0.7397 & 0.7312 & 0.8365 & 0.7789 & 0.7358 & 0.7523 \\
            
            \midrule
            \midrule 
            
            \textbf{Traffic-MoE(Ours)} & \textbf{0.9089} & \underline{0.8942} & \textbf{0.7879} & \underline{0.8072} & \textbf{0.7613} & \textbf{0.7866} & \textbf{0.8158} & \textbf{0.8005} & \textbf{0.8588} & \underline{0.8007} & \underline{0.7701} & \underline{0.7824} \\
            w/ Header Only & 0.8759 & 0.7721 & 0.7106 & 0.6996 & 0.7308 & 0.7665 & 0.7846 & 0.7715 & 0.8395 & 0.7679 & 0.7211 & 0.7358 \\
            w/ Payload Only & 0.6227 & 0.6397 & 0.5566 & 0.5696 & 0.4967 & 0.6236 & 0.5445 & 0.5472 & 0.3965 & 0.5782 & 0.3314 & 0.3710 \\
            w/o Payload-Data & \underline{0.9068} & \textbf{0.8944} & \underline{0.7798} & \textbf{0.8096} & \underline{0.7610} & \underline{0.7856} & \underline{0.8097} & \underline{0.7931} & \underline{0.8583} & \textbf{0.8045} & \textbf{0.7713} & \textbf{0.7871} \\
            w/o Payload-Negotiation & 0.7435 & 0.7095 & 0.6073 & 0.6242 & 0.6217 & 0.6758 & 0.6381 & 0.6087 & 0.7022 & 0.6982 & 0.6163 & 0.6389 \\
            w/o Header-Length & 0.7975 & 0.8618 & 0.5973 & 0.6004 & 0.6743 & 0.6383 & 0.6677 & 0.6478 & 0.5235 & 0.6155 & 0.5178 & 0.4636 \\
            w/o Header-Direction & 0.8937 & 0.8912 & 0.7713 & 0.7961 & 0.7349 & 0.7449 & 0.7717 & 0.7572 & 0.6747 & 0.7095 & 0.6492 & 0.6199 \\
            w/o Header-Flags & 0.8684 & 0.8910 & 0.6870 & 0.7204 & 0.7015 & 0.7371 & 0.7263 & 0.7311 & 0.6504 & 0.7152 & 0.6244 & 0.6219 \\
            w/o Header-Proto & 0.9013 & 0.8937 & 0.7702 & 0.7844 & 0.7513 & 0.7768 & 0.7945 & 0.7848 & 0.7928 & 0.7157 & 0.6794 & 0.6781 \\
            w/o Header-Time & 0.7595 & 0.6618 & 0.5808 & 0.5938 & 0.5054 & 0.6344 & 0.5943 & 0.5578 & 0.5521 & 0.6436 & 0.4282 & 0.4407 \\
            
            \bottomrule
        \end{tabular}
    }
\end{table*}

\subsection{Hyperparameter Analysis of the Top-$k$ Routing Strategy}
To assess the trade-off between representation capacity and computational sparsity, we evaluate varying the number of active experts, $k \in \{1, 2, 3, 4, 6\}$ (Figure~\ref{fig:impact_topk}). Results show that across most scenarios, performance peaks at $k=2$, achieving Pareto optimality between accuracy and inference efficiency.
\begin{figure}[!t]
    \centering
    \includegraphics[width=\linewidth]{FigAPP-10-topk-r1c4.pdf}
    \caption{Impact of Top-$k$ parameters on model performance.}
    \label{fig:impact_topk}
\end{figure}

Specifically, increasing $k$ from 1 to 2 yields significant gains, primarily attributed to collaborative synergy between "primary-auxiliary" experts in feature extraction. However, as $k$ further increases, the model exhibits performance fluctuation and degradation across all scenarios (\eg \textit{ISCXVPN2016}). This performance decay reveals the core mechanism of the MoE architecture: its advantages stem from sparse activation and specialized isolation rather than global computation. Under the design constraint of a constant active parameter count, forcing the activation of more experts leads to a severe dilution of parameter capacity assigned to each individual expert, thereby weakening their representational power.
Crucially, a larger $k$ forces heterogeneous traffic features to share the same expert subsets, breaking the established state of specialized isolation. This triggers significant feature interference and blurs the classification boundaries. In extreme cases, an excessive increase in $k$ causes the MoE architecture to degenerate into a structure resembling traditional dense networks, significantly undermining the model's efficacy in processing complex, heterogeneous network traffic.

\subsection{Parameter Analysis on Traffic Sequence}
To explore the optimal sequence truncation strategy and balance representational capacity with computational overhead, we conducted an ablation study on the payload truncation length ($J$) and the number of truncated packets ($K$).

\textbf{Impact of Payload Truncation Length ($J$):} As shown in Figure~\ref{fig:payload}, the model's performance exhibits an overall upward trend as $J$ increases from 10 to 40 bytes, after which it plateaus. This phenomenon validates the inherent \textit{``header semantic concentration''} characteristic of network traffic: critical discriminative fingerprints (\eg TLS record headers and cipher suite identifiers) are highly clustered at the beginning of the payload. Regions beyond this threshold mostly consist of high-entropy encrypted data or random padding, which lack discriminative power and act as noise that disrupts the attention mechanism's focus. Therefore, setting $J=40$ precisely extracts high-density discriminative semantics while effectively bounding the token sequence length, drastically reducing FLOPs during inference.

\textbf{Impact of Packet Count ($K$):} As Figure~\ref{fig:packet} illustrates, the majority of task scenarios exhibit a clear pattern of diminishing marginal returns. Performance significantly improves as $K$ increases up to 10; however, for $K > 10$, it generally plateaus or fluctuates. This occurs because the most discriminative ``protocol syntax'' (\eg TCP/TLS handshakes) manifests entirely during the initial connection establishment phase. Once entering the bulk data transfer phase, traffic behavior becomes highly homogenized, offering negligible incremental information. Thus, for most tasks, $K=10$ is the optimal solution to fully capture session logic while preventing the quadratic growth of computational complexity.

Notably, in the modern pure TLS 1.3 encrypted scenario (\textit{CipherSpectrum}), expanding the observation window yields continuous gains. Although the model already demonstrates excellent discriminative capability at $K=10$, increasing $K$ to 20 achieves further breakthroughs. This reflects a TLS 1.3 representational shift: because comprehensive encryption drastically compresses plaintext handshake phase, more discriminative features are implicitly encoded within mid-to-late interactions (\eg application-level packet length evolution and temporal burst). Consequently, longer observation window effectively unlocks these deep behavioral features, supplying abundant structural cues to sharpen the decision boundary.

\begin{figure}[!t]
    \centering
    \includegraphics[width=\linewidth]{FigAPP-11-payload.pdf}
    \caption{Impact of payload length on model performance.}
    \label{fig:payload}
\end{figure}

\begin{figure}[!t]
    \centering
    \includegraphics[width=\linewidth]{FigAPP-12-packnum.pdf}
    \caption{Impact of packet count on model performance.}
    \label{fig:packet}
\end{figure}


\subsection{Detailed Evaluation under Distribution Shifts}\label{app: ood}
\begin{itemize}[leftmargin=10pt]
    \item \textbf{Time-shift} simulates concept drift. For fine-grained traffic classification tasks (\eg distinguish DDoS sub-types), we strictly sort all session flows of each category by timestamp. We designate the first 40\% of the time span as the training domain and the last 40\% as the testing domain, discarding the middle 20\% to establish a temporal buffer. This ``past predicting future'' setting eliminates temporal data leakage and ensures a significant temporal distributional gap.
    \item \textbf{Proportion-shift} evaluates the model's adaptability to intra-class distribution shifts. For coarse-grained traffic classification tasks, we artificially adjust the ratio of dominant to minor components in each coarse-grained category (\eg DDoS, WebAttack): the training set is sampled with a 4:1 ratio, whereas the testing set adopts a completely inverse 1:4 ratio. This extreme distribution inversion rigorously tests the model's capacity to capture long-tail features.
    \item \textbf{Compose-shift} simulates unseen attack sub-variants. In coarse-grained traffic classification tasks, we randomly mask 50\% of fine-grained sub-class traffic per category during training, while the testing set includes all sub-classes, thereby simulating the real-world challenge of ``missing information'' due to emerging attack variants.
\end{itemize}

\subsubsection{Resilience to Concept Drift (Time-shift)}
The severe performance degradation of dense baselines like ET-BERT under the Time-shift scenario exposes their structural reliance on \textit{temporal shortcuts}. These shortcuts include transient statistical artifacts tied to the training timeframe, which inevitably become invalid as the network context evolves. Conversely, the sparse expert specialization of \model naturally resists this overfitting. By selectively activating experts grounded in pre-trained protocol knowledge, the architecture inherently filters out temporal noise and isolates invariant attack behaviors, maintaining high robustness against concept drift.

\subsubsection{Mitigation of Gradient Starvation (Proportion-shift)}
The competitive performance under intra-class distribution shifts shows our MoE architecture effectively mitigates the gradient starvation of minority classes typically encountered in dense networks. The routing mechanism acts as a divide-and-conquer strategy, assigning long-tail minor components to specific experts. This prevents the gradient updates of these minority variants from being diluted by dominant classes, ensuring high recall even under inverted test distributions.

\subsubsection{Adaptation to Unseen Variants (Compose-shift)} In scenarios containing unseen sub-classes, the consistently high recall of \model further validates its capacity for behavioral abstraction. Instead of overfitting to tool-specific payloads, the model captures generic malicious patterns (\eg structural characteristics of flooding attack). This ability to generalize rather than simply memorize signatures is critical for identifying emerging threats, particularly in operational environments where evasions carry severe security costs.

By effectively managing the inherent non-stationarity of real-world traffic, \model avoids catastrophic performance drops common in dense models. Its superior generalization enables sustained adaptation to evolving network environments, significantly reducing the frequency and computational overhead of model retraining in long-term deployments.

\subsection{Empirical Justification for Dense Attention Mechanism}\label{app:dense_attention}
\begin{figure}[!t]
    \centering
    \includegraphics[width=\linewidth]{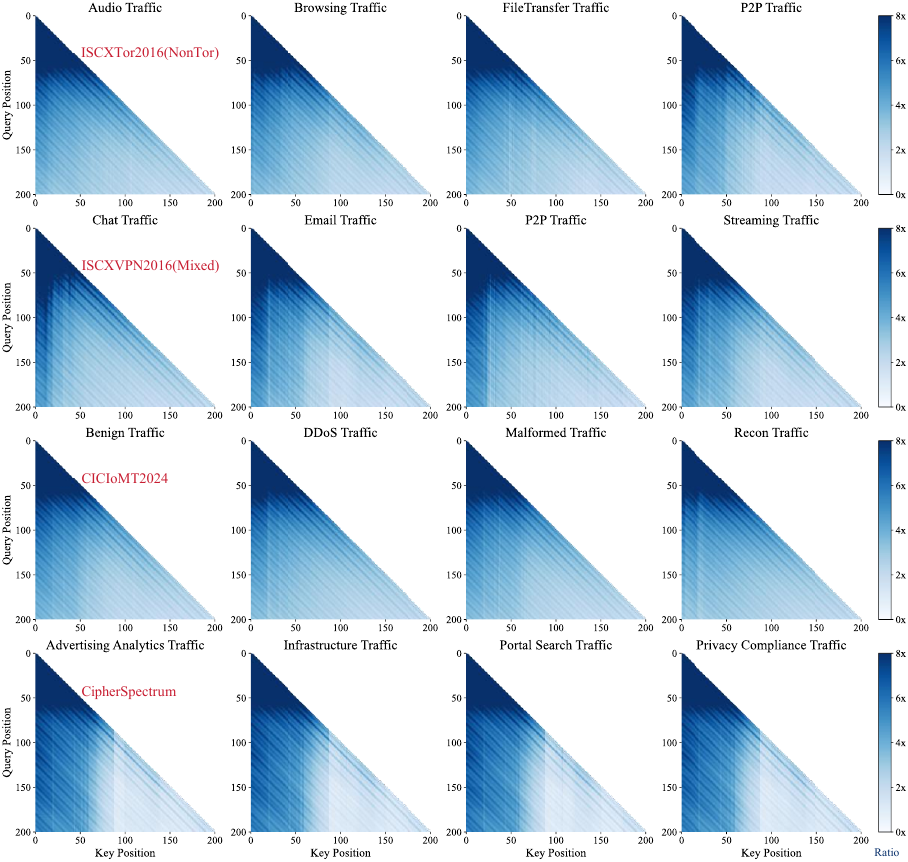}
    \caption{Attention distributions of traffic categories across detection tasks.}
    \label{fig:attn-weight}
\end{figure}

To empirically justify the design choice of employing dense attention in \model for capturing universal traffic syntax, we visualize the attention distributions of representative traffic categories across four detection scenarios (see Figure~\ref{fig:attn-weight}). We average the multi-head attention matrices at specific model layers over the entire test set and apply padding masks to ensure the computation focuses on valid traffic tokens. For maximum information density, the visualization focuses on the first half of flow sequences and utilizes the ``Attention-to-Baseline Ratio'' as a quantitative metric, where $1\times$ represents the uniform attention baseline.

\textbf{Capturing Universal Protocol Syntax.} Despite significant semantic divergence among different traffic categories, their attention patterns exhibit striking structural consistency at the same model depth. This indicates that the dense attention module effectively functions as a \textit{universal grammars parser}. Since the vast majority of network traffic (whether benign business flows or malicious attacks) strictly adheres to foundational TCP/IP state machines and structural delimiters, the attention mechanism successfully extracts cross-class shared representations rather than overfitting to class-specific signatures.

\textbf{Interpretation of Attention Patterns.} The observed attention distributions align with the physical properties of network protocols. First, we observe a pronounced \textit{header focus}. Driven by causal mask dilution and the high information entropy inherent in initial flow negotiations and metadata exchanges, attention concentrates at the flow sequence start. Local attention peaks in this region reach up to $8\times$ the baseline, quantitatively validating the backbone's precision in parsing foundational syntax. Second, the distinct diagonal stripes in the heatmaps reveal the sequential parsing of local byte-level transitions. This localized context extraction naturally complements the global long-range dependencies captured by the header focus, collectively enabling the model to construct comprehensive hierarchical representations.


\textbf{Strategic Architectural Choice.} Since network traffic syntax remains invariant across distinct categories, imposing artificial sparsity constraints on the attention mechanism would introduce unnecessary inductive bias, resulting in a suboptimal architecture. Our empirical findings substantiate that the functionally decoupled design (\textit{dense attention for universal syntax extraction, paired with sparse MoE for specialized semantic modeling}) achieves optimal representational capacity with minimal computational redundancy. This architectural synergy fundamentally explains how \model sustains robust generalization while maintaining highly efficient inference.

